\documentclass[prb,twocolumn,aps,superscriptaddress,showpacs]{revtex4}

\usepackage{epsfig}
\usepackage{graphicx}
\usepackage{verbatim}
\usepackage{comment}
\usepackage{amsmath}
\usepackage{amssymb}
\usepackage{stmaryrd}
\usepackage{color}
\usepackage{calrsfs}
\DeclareMathAlphabet{\pazocal}{OMS}{zplm}{m}{n}
\usepackage{epstopdf}

\begin{document}

\title{Finite-temperature properties of strongly correlated fermions in the honeycomb lattice}

\author{Baoming Tang}
\affiliation{Department of Physics, The Pennsylvania State University,
University Park, Pennsylvania 16802, USA}
\affiliation{Department of Physics, Georgetown University, Washington
DC, 20057 USA}
\author{Thereza Paiva}
\affiliation{Instituto de F\' isica, Universidade Federal do Rio de Janeiro 
Cx.P. 68.528, 21941-972 Rio de Janeiro RJ, Brazil}
\author{Ehsan Khatami}
\affiliation{Physics Department, University of California, Santa Cruz, California 95064, USA}
\author{Marcos Rigol}
\affiliation{Department of Physics, The Pennsylvania State University,
University Park, Pennsylvania 16802, USA}

\begin{abstract}
We study finite-temperature properties of strongly interacting fermions in the honeycomb 
lattice using numerical linked-cluster expansions and determinantal quantum Monte 
Carlo simulations. We analyze a number of thermodynamic quantities, including the entropy, 
the specific heat, uniform and staggered spin susceptibilities, short-range spin 
correlations, and the double occupancy at and away from half filling. We examine the 
viability of adiabatic cooling by increasing the interaction strength for homogeneous 
as well as for trapped systems. For the homogeneous case, this process is found to be 
more efficient at finite doping than at half filling. That, in turn, leads to an efficient 
adiabatic cooling in the presence of a trap which, starting with even relatively high 
entropies, can drive the system to have a Mott insulating phase with substantial 
antiferromagnetic correlations.
\end{abstract}

\maketitle
                                                                                                                                          
\section{INTRODUCTION}

The Fermi-Hubbard model in the honeycomb lattice, especially its ground-state phase 
diagram, has attracted much interest in recent years. This is in part motivated by the 
advent of graphene,\cite{castroneto09} whose semimetallic properties can be understood within 
the weakly interacting regime of this model. It is also motivated by results 
of large-scale quantum Monte Carlo (QMC) calculations by Meng {\it et al.},\cite{meng_lang_10} 
which suggested that a spin liquid phase exists in that model at intermediate 
interaction strengths, despite the absence of frustration. Following those results, a considerable 
number of studies have been done on the Hubbard model as well as in closely related 
models in the honeycomb geometry.\cite{w_wu_10,a_liebsch_11,m_ebrahimkhas_11,t_ma_11,g_wang_11,
g_sun_11,Tang_12_01,r_he_12,c_chang_12,Sandro_Yuichi_12,w_wu_13,s_hassan_13,assaad_herbut} 
However, QMC simulations of very large systems,\cite{Sandro_Yuichi_12} and the
use of accurate approaches to estimate the antiferromagnetic (AF) order parameter,\cite{assaad_herbut} 
have indicated that the existence of a spin liquid in such a model is unlikely.

A remarkable property of the Fermi-Hubbard model in the honeycomb 
lattice at half filling is the existence of a quantum phase 
transition to a Mott insulator at a finite interaction strength. Insulating behavior 
in this case is accompanied by the emergence of long-range AF correlations 
at zero temperature, as unveiled in QMC simulations.\cite{s_sorella_92,Paiva_05,Sandro_Yuichi_12} 
This phase transition has also been examined by means of dynamical mean-field theory and its 
cluster extensions,\cite{g_santoro_93,s_jafari_09,w_wu_10,a_liebsch_11,m_ebrahimkhas_11,r_he_12,
s_hassan_13} as well as within the coherent phase approximation.\cite{d_le_13} Other studies 
of this and closely related models have explored the existence of pairing and superconductivity 
away from half filling.\cite{t_ma_11,w_wu_13}

A new front for studying the properties of the Fermi-Hubbard model in the honeycomb lattice
has been recently opened by the (almost ideal) experimental realization of this model 
by: (i) building artificial graphene and loading it with a two-dimensional electron 
gas,\cite{honeycom_sci} (ii) trapping ultracold atoms in highly tunable optical lattices 
with that geometry,\cite{honeycom_nat1} and (iii) creating molecular graphene in a 
fully tunable condensed-matter set up.\cite{honeycom_nat2} These experiments follow 
extensive research on simulating the Hubbard model on cubic geometries using ultracold
gases,\cite{esslinger_review_10} aimed at answering fundamental questions surrounding 
Hubbard models, such as whether they support superconductivity.

In a recent Letter,\cite{Tang_12_01} we utilized two state-of-the-art computational 
methods, a numerical linked-cluster expansion (NLCE) and determinantal quantum Monte 
Carlo (DQMC), to study thermodynamic properties and spin correlations of 
interacting fermions in the honeycomb lattice, as described by the Fermi-Hubbard 
model. Our emphasis was in the half-filled case for which we found that, in comparison to 
the square lattice, the honeycomb lattice exhibits a more significant region of anomalous 
rise in the double occupancy at low temperature, leading to a more efficient adiabatic 
cooling. We also showed that nearest-neighbor (NN) spin correlations are stronger in 
the honeycomb lattice than in the square lattice for a wide range of interaction strengths
and for an extended region of experimentally relevant entropies. This was found to be 
true even in the weakly-interacting regime, where the former has a semimetallic ground state
while the latter has an AF Mott insulating one. 

In this work, we extend the analysis presented in Ref.~\onlinecite{Tang_12_01} 
by studying other quantities of much interest, such as the entropy, the specific heat, 
uniform and staggered spin susceptibilities, and the structure factor. We 
establish a connection between the temperature at which the nearest neighbor (NN) 
spin correlations change most rapidly and the onset of exponentially long AF 
correlations in the system. More importantly, 
we explore in detail the behavior of some of those quantities, and of quantities studied 
in Ref.~\onlinecite{Tang_12_01}, away from half filling. In particular, we analyze 
how the double occupancy, which can be accurately measured in ultracold gases 
experiments,\cite{jordens_strohmaier_08} evolves as a function of temperature at 
different densities. Isentropic curves for the temperature as a function of the 
interaction strength away from half filling are also presented. Finally, employing a local
density approximation (LDA), we analyze the possibility of achieving adiabatic cooling by 
increasing the strength of the interaction between trapped fermions in optical lattice 
experiments. We show that in the presence of a harmonic trap, even starting with relatively 
high entropies, it is possible to create Mott insulating regions with exponentially long 
antiferromagnetic correlations by using this approach. 

The exposition is organized as follows. In Sec.~\ref{sec:model}, we introduce the Hubbard 
model and provide a brief explanation of the two numerical methods used in this study. In
Sec.~\ref{sec:results}, we report results for the thermodynamic properties of the homogeneous
system in the honeycomb lattice and, in some cases, we compare those results with the ones 
obtained in the square lattice. Properties of systems confined by a harmonic potential,
relevant to ultracold gases experiments, are presented in Sec.~\ref{sec:trapped}. A summary 
of our results is presented in Sec.~\ref{sec:summ}.

\section{MODEL AND COMPUTATIONAL METHODS}
\label{sec:model}

\subsection{The Hamiltonian}

The one-band Hubbard Hamiltonian describes electrons in a lattice, or two 
component fermions in an optical lattice, and can be written as
\begin{equation} \label{eq:Hamiltonian}
 \hat{H}=-t\sum_{\langle i,j \rangle \sigma}(\hat{c}^\dagger_{i\sigma}\hat{c}^{}_{j\sigma}+ 
 \text{H.c.}) + U\sum_{i}\hat{n}^{}_{i\uparrow}\hat{n}^{}_{i\downarrow},
\end{equation}
where $\hat{c}^\dagger_{i\sigma}$ ($\hat{c}^{}_{i\sigma}$) is a creation (annihilation) 
operator of a fermion with spin (or pseudo-spin) $\sigma$ on lattice site $i$, and 
$\hat{n}_{i\sigma}=\hat{c}^\dagger_{i\sigma} \hat{c}^{}_{i\sigma}$ is the site occupation 
operator. Here, $\langle ..\rangle$ denotes nearest-neighbor sites, $t$ is the 
nearest-neighbor hopping amplitude, and $U>0$ is the on-site repulsive interaction.

The most recent DQMC studies of the phase diagram of this model in the honeycomb 
geometry indicate that there are two ground-state phases, a semimetallic one at weak 
coupling and an AF Mott insulating one at strong coupling, with a continuous phase 
transition between them that occurs at $U_{c}/t\simeq 3.8$.\cite{Sandro_Yuichi_12,assaad_herbut}

\subsection{Numerical approaches}

In this work, we use two fundamentally different unbiased numerical methods, a
NLCE\cite{Marcos_06,Marcos_07_01,Marcos_07_02} and DQMC,\cite{Scalapino_81}
to study the thermodynamic properties of Hamiltonian \eqref{eq:Hamiltonian} in the 
honeycomb lattice.

In NLCEs, any extensive property of a lattice model per site $P(\mathcal{L})/N$
($N$ is the number of lattice sites), {\em in the
thermodynamic limit}, can be expanded in terms of contributions from all clusters
$c$ that can be embedded in the infinite lattice:
\begin{equation}
\label{NLCE1}
P(\mathcal{L})/N=\sum_{c}L(c)\times W_{P}(c),
\end{equation}
where $L(c)$ is the lattice constant of $c$, defined as the number of ways per site in which
cluster $c$ can be embedded in the lattice, and $W_{P}(c)$ is the weight of that cluster.
$W_{P}(c)$ can be computed recursively based on the inclusion-exclusion principle:
\begin{equation}
\label{NLCE2}
 W_{P}(c)=P(c)-\sum_{s \subset c} W_{P}(s),
\end{equation}
where $P(c)$ is the property calculated for the finite cluster $c$ using full exact
diagonalization.\cite{Marcos_06,Marcos_07_01,Marcos_07_02}

In this work, we use the site-based NLCE,\cite{Marcos_07_01} i.e., contributions to the
$m\textrm{th}$ order of the expansion come solely from clusters with up to $m$ sites.
The computational effort 
is reduced by identifying all clusters that have the same Hamiltonian (same topology), and 
diagonalizing only one of them.\cite{Marcos_07_01,Tang_12} In Table I, we report the total 
number of topologically distinct clusters (second column) that are diagonalized in 
each order of the NLCE (first column). The number of topological clusters 
should be compared to the much larger number of added lattice constants at each order 
of the expansion (third column).
\begin{table}[b]
\caption{Number of topologically distinct clusters and sum of the lattice constants
for clusters with up to 17 sites in the honeycomb lattice.}
\begin{tabular}{rrr}
\hline\hline
No.\ of sites & No.\ of topological clusters & $\sum L(c)$ \\
\hline 
   1  &       1  &\qquad               1 \\
   2  &       1  &\qquad             3/2 \\
   3  &       1  &\qquad               3 \\
   4  &       2  &\qquad               7 \\
   5  &       2  &\qquad              18 \\
   6  &       5  &\qquad              47 \\
   7  &       7  &\qquad             125 \\
   8  &      15  &\qquad           675/2 \\
   9  &      26  &\qquad             919 \\
  10  &      59  &\qquad        5\,053/2 \\
  11  &     113  &\qquad          7\,008 \\
  12  &     258  &\qquad       39\,169/2 \\
  13  &     542  &\qquad         55\,097 \\
  14  &  1\,233  &\qquad      311\,751/2 \\
  15  &  2\,712  &\qquad        443\,080 \\ 
  16  &  6\,208  &\qquad     1\,264\,630 \\
  17  & 14\,004  &\qquad     3\,622\,431 \\
\hline\hline
\end{tabular}
\label{bondclusters}
\end{table}

NLCEs converge at a given temperature whenever correlations in the system are of the 
order of or smaller than the largest cluster sizes considered in the expansion. However, 
even if the bare sums diverge, one can accelerate the convergence of NLCEs at lower 
temperatures by using resummation algorithms. Here, we carry out the calculations up 
to the ninth order in the site expansion and employ Wynn and Euler resummation
techniques\cite{Marcos_07_01,Tang_12} to improve the convergence at 
low temperatures. Our calculations are performed in the grand-canonical
ensemble,\cite{Marcos_07_02} and we compute all observables for a wide range 
(and dense grid) of chemical potentials, $\mu$, and temperatures, $T$. Note that we 
can make the grid for $\mu$ and $T$ arbitrarily fine without much computational overhead 
since, for a given $U$, everything needed to compute the observables is generated in a single
run of the exact diagonalization. This allows us to accurately compute the equation of 
state so that all quantities can also be studied at any desired constant density.

Within DQMC,\cite{Scalapino_81} the partition function is expressed as a path 
integral by using the Suzuki-Trotter decomposition of $\exp(-\beta \hat{H})$, where 
$\beta=1/ (k_B T)$ with $k_B$ the Boltzmann constant, after discretizing the imaginary time. 
This introduces an imaginary-time 
interval $\Delta\tau$ that we set to $\Delta \tau \times t=0.05$. The interaction 
term is decoupled through a discrete Hubbard-Stratonovich transformation,\cite{hirsch_83} 
which introduces an auxiliary Ising field. One can then integrate the fermionic 
degrees of freedom analytically, and the summation over the auxiliary field (which 
depends both on the site and the imaginary time) is carried out stochastically using a 
Monte Carlo algorithm. 
Most of our results are for 96 sites for the honeycomb lattice and 100 sites for the 
square lattice geometry, though in some cases, other system sizes (namely 24, 54, 150, and 
216 sites) were studied for the honeycomb lattice to analyze finite-size effects. As it 
will become apparent in the following, from comparisons to the NLCE, the systematic errors 
in DQMC are mostly negligible for our parameters of interest.

As shown in previous studies,\cite{Ehsan_Marcos_11,Tang_12_01} the two approaches are
complementary. For the half-filled system with weak interactions (in comparison
to the bandwidth), DQMC can access lower temperatures than NLCE. This can be
understood from the fact that the computational cost of the former method increases almost
linearly with inverse temperature, whereas the exponential growth of correlations in the
model by decreasing the temperature limits the region of convergence of NLCE. 
On the other hand, for strong interactions, while DQMC runs 
into statistical difficulties when sampling the auxiliary fields, the convergence of 
NLCE extends to lower temperatures. This is because in the strong-coupling regime of the 
half-filled system, the relevant energy scale at low temperatures is that of the spin degrees 
of freedom, namely, the exchange interaction $J\propto 1/U$, which sets the characteristic 
temperature for the onset of the exponential growth of AF correlations. This scale decreases 
as the interaction strength increases and NLCEs can follow it without running into 
computational difficulties.

Away from half filling, DQMC is limited to higher temperatures in comparison to half filling 
due to the so-called fermion `sign problem'. \cite{sign1,sign2} We find that in NLCEs, despite 
the lack of sign problem, the lowest temperature at which the series converge for our observables
of interest also generally increases as the system is doped away from half filling. This 
seems to be a consequence of the emergence of competing correlations at higher temperatures.

\section{RESULTS}
\label{sec:results}

\subsection{Entropy}

In the grand-canonical ensemble, the entropy (per site) can be written as
\begin{equation}\label{entropy}
S=\frac{\ln Z}{N} + \frac{\langle\hat{H}\rangle - \mu \langle{\hat{N}_p}\rangle}{N\,T},
\end{equation} 
where $\hat{N}_p$ is the operator for the total number of particles.
Within NLCEs, all the extensive quantities in the right-hand side of Eq.~\eqref{entropy} 
can be computed (per site) directly in the thermodynamic limit. This is not the case within 
DQMC. The calculation of the entropy in the latter approach, for finite clusters, 
is done by integrating the energy\cite{Werner_05}
\begin{equation}
S(T) = S(T \to \infty) +  \frac{\beta \langle\hat{H}\rangle}{N}
- \frac{1}{N}\int_0^\beta \langle\hat{H}\rangle\, d\beta,
\end{equation} 
where $S(T \to \infty)$ is the high-temperature limit of the entropy for a given 
density $n$; $S(T \to \infty) = 2\ln (2) - n \ln (n) - (2-n) \ln (2-n)$. Throughout 
the paper, the entropy is given in units of $k_B$. Since we present 
results for two different lattice geometries, we have chosen the unit of energy to be 
the noninteracting bandwidth, $w$, which is $6t$ for the honeycomb lattice and $8t$ for
the square lattice.

\begin{figure}[!t] 
\includegraphics[width=0.46\textwidth]{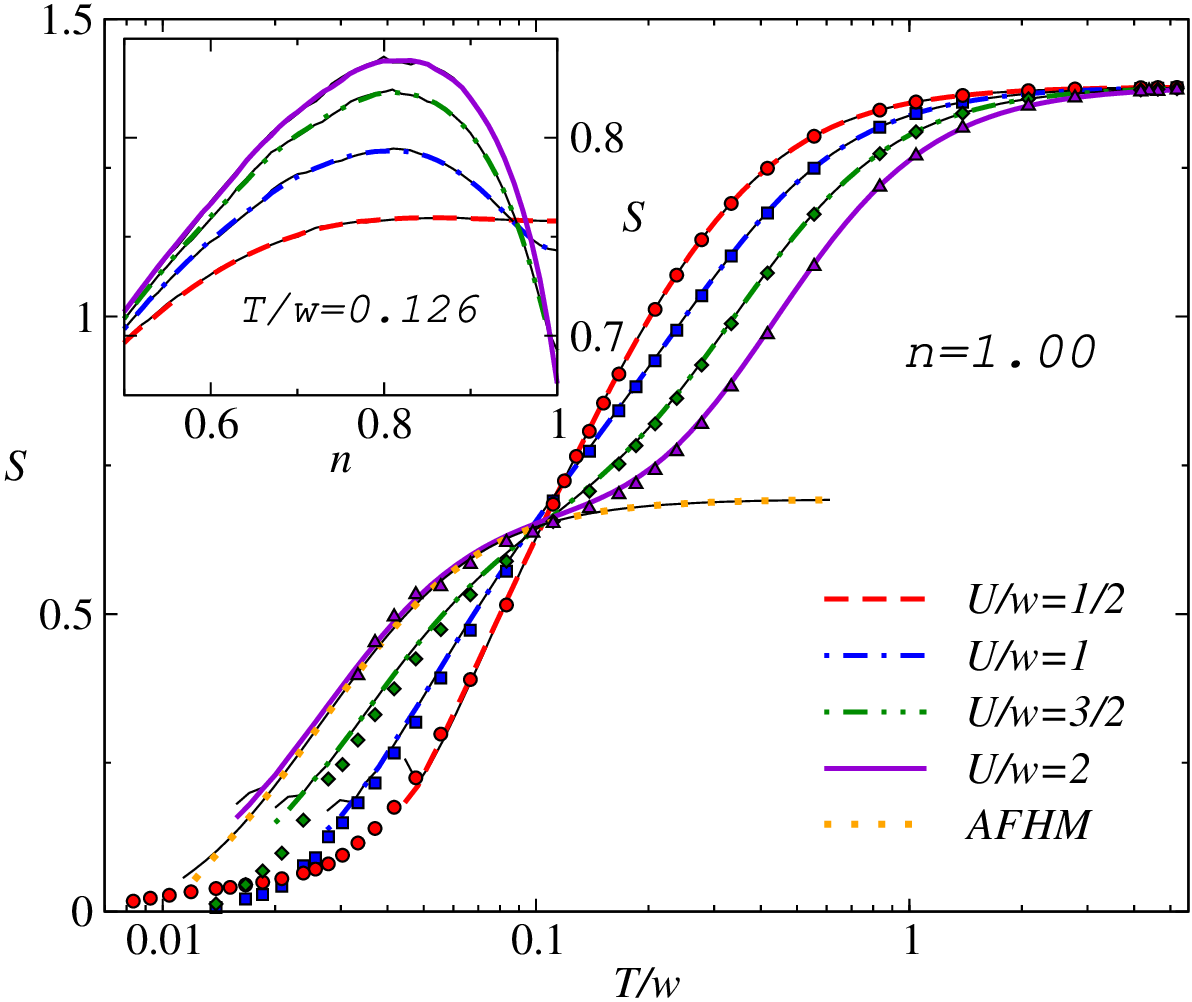}
\caption{(Color online) NLCE (lines) and DQMC (symbols) results for the entropy of 
the Hubbard model in the honeycomb (HC) lattice as a function of $T$ for $U/w=1/2$, 
$1$, $3/2$, and $2$ at half filling. The NLCE results were obtained by applying 
Euler's resummation to the last six terms in each order of the 
expansion.\cite{Marcos_07_01,Tang_12} Here we report results for the last order 
(thick lines) and the next to last order (thin black lines). We also show NLCE 
results for the entropy of the antiferromagnetic Heisenberg model (AFHM) for 
$J/w = 1/18$. The site expansion for this model was carried out up to the 
seventeenth order. The results in this case were obtained using Wynn's 
algorithm and we report those for the highest term after eight (thick line) 
and seven (thin black line) orders of improvement.\cite{Marcos_07_01,Tang_12} Unless 
otherwise specified, all NLCE results reported in the following for the honeycomb
lattice are obtained as explained above and are presented using the same convention. 
Inset: NLCE results for the entropy vs density ($n$) at 
$T/w=0.126$ for the same values of $U$ and legends used in the main panel.}
\label{fig:S_Cv_vs_T}
\end{figure}

In the main panel in Fig.~\ref{fig:S_Cv_vs_T}, we show the entropy of the honeycomb 
lattice as a function of temperature at half filling and for different values of $U$ 
from NLCE (lines) and DQMC (symbols). As anticipated, we find that NLCE results converge 
to lower temperature as $U$ increases. For the largest value of $U$ shown ($U/w=2$), 
the convergence is extended to around $T/w=0.02$. 
Figure~\ref{fig:S_Cv_vs_T} shows that, for the two smallest values 
of $U$ ($U/w=1/2$ and 1), the results from the two methods agree very well down to the 
lowest convergence temperature in the NLCE. On the other hand, small deviations between 
the results from NLCE and DQMC become apparent for larger values of $U$ and low 
temperature, e.g., DQMC slightly underestimates the entropy for $U/w=1.5$ for $T/w<0.1$. 

As expected, and similarly to previous results for the square lattice,\cite{Ehsan_Marcos_11} 
the entropy shows different behavior in the weak- and strong-coupling regimes. While 
in the former it decreases steadily from $\ln4$ to $0$ when lowering the temperature, 
in the latter, a plateau develops at intermediate temperatures ($T/w \sim0.1$). The plateau 
becomes visible with increasing $U$. This can be explained as follows: when
$U$ (and consequently the cost of double occupancy) increases the charge degrees of 
freedom freeze out at increasingly high temperatures. Once they are frozen, the system 
at lower temperatures can be described by an effective Heisenberg model (with  
$J=4t^2/U$), whose high-temperature entropy ($S\sim\ln2$) agrees with the entropy at 
the plateau of the Hubbard model. As $U$ increases, $J$ decreases and the low temperature 
regime of the effective Heisenberg model moves to lower temperatures. This results in a larger
plateau with increasing $U$. To make those points apparent, in Fig.~\ref{fig:S_Cv_vs_T} we 
also show the entropy of the AF Heisenberg model in the honeycomb lattice with 
$J/w = 1/18$, which closely follows the results of the Hubbard model with $U/w=2$ at $T/w <0.1$.

NLCE results for the entropy away from half filling are presented in the inset in
Fig.~\ref{fig:S_Cv_vs_T}. Those results were obtained for $T/w=0.126$, the same four 
values of $U$ depicted in the main panel, and for a wide range of local fillings 
($0.5\leq n\leq 1$). In the weakly-interacting regime (e.g., $U/w=1/2$), 
as the density increases, the entropy increases monotonically first and then saturates 
at a finite value when $n > 0.8$. This is because the system behaves similarly 
to a non-interacting system at this temperature: as $n\to1$, the number of microstates increases
and therefore the entropy increases. However, as correlations become significant 
by increasing $U$, the entropy decreases dramatically as the density approaches 
half filling due to the formation and ordering of local moments. This is 
accompanied by the appearance of a peak in the entropy at $n\sim 0.85$, which 
resembles the one appearing for the same model in the square 
lattice.\cite{Bon_03,Mikelsons_09,Ehsan_Marcos_11} The high 
entropy around that filling at low temperature signals the presence of many
low-energy competing states, which, upon further cooling, could result in the 
emergence of exotic phases such as superconductivity.

\subsection{Specific Heat}

The specific heat (per site) is defined as 
\begin{eqnarray}\label{eq:heat0}
C_v&=&\frac{1}{N} \left( \frac{\partial\langle\hat{H}\rangle}{\partial T} \right)_{n} \\
   &=&\frac{1}{N}\left[ \left( \frac{\partial\langle\hat{H}\rangle}{\partial T} \right)_{\mu} + 
     \left(\frac{\partial\langle\hat{H}\rangle}{\partial \mu} \right)_{T}
     \left(\frac{\partial\mu}{\partial T} \right)_{n}\right], \label{eq:heat}
\end{eqnarray}
where the second expression is more amenable for evaluation in the grand-canonical 
ensemble. However, in order to avoid numerical derivatives and eliminate the systematic 
errors they introduce, we rewrite Eq.~\eqref{eq:heat} using Maxwell equations 
as\cite{Ehsan_Marcos_12}
\begin{equation} \label{eq:heat_01}
 C_v= \frac{1}{NT^2}\left[ \langle\Delta\hat{H}^2\rangle - \frac{\left(\langle\hat{H}
\hat{n}\rangle - \langle\hat{H}\rangle \langle\hat{n}\rangle \right)^2}{\langle \Delta
\hat{n}^2 \rangle} \right], 
\end{equation}
where $\langle\Delta\hat{H}^2\rangle = \langle\hat{H}^2\rangle - \langle\hat{H}\rangle^2$ 
and $\langle\Delta\hat{n}^2\rangle = \langle\hat{n}^2\rangle - \langle\hat{n}\rangle^2$. 
All expectation values in Eq.~\eqref{eq:heat_01} can be directly evaluated in NLCEs.

Within DQMC, obtaining $C_v$ from Eq.~\eqref{eq:heat_01} requires the calculation of 
expectation values of products of up to eight fermion operators. While this is possible 
in principle, it would lead to large statistical fluctuations in the results, 
which are costly to reduce. Therefore, we resort to the numerical 
differentiation of high-quality data for $\langle\hat{H}\rangle$ as prescribed 
in Eq.~\eqref{eq:heat0}.

\begin{figure}[!t] 
\includegraphics[width=0.44\textwidth]{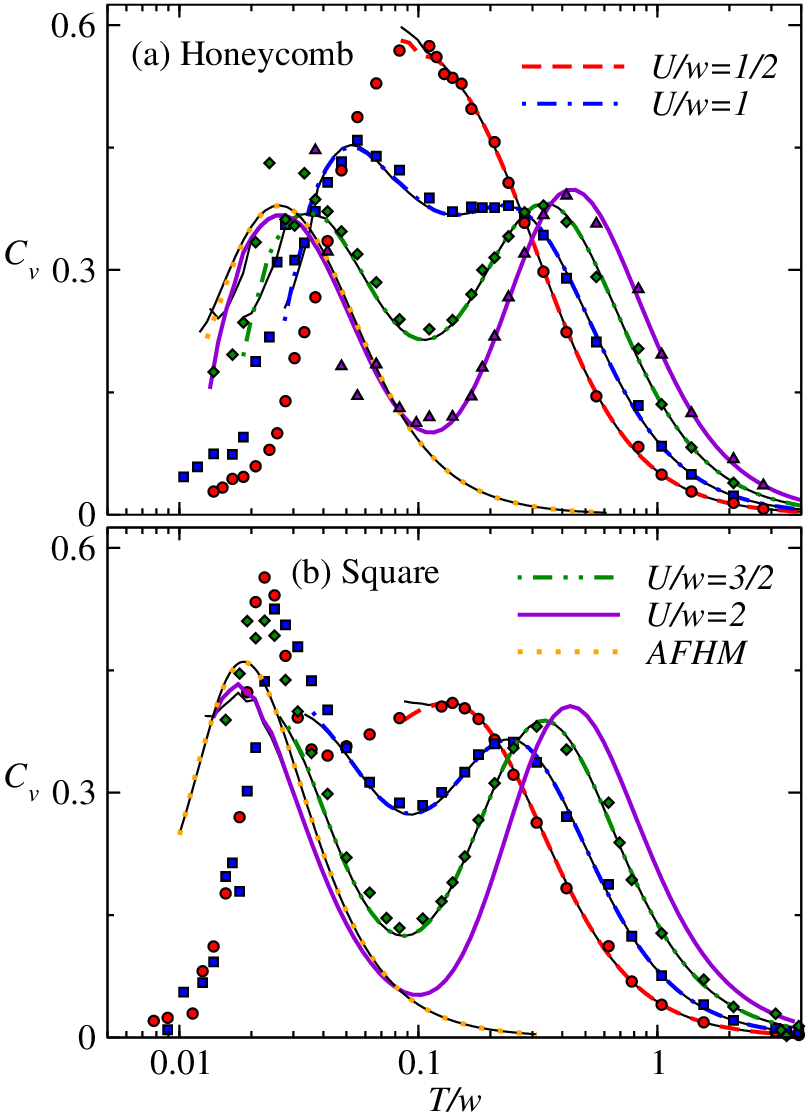}
\caption{(Color online) Specific heat vs $T$ for (a) the honeycomb lattice and (b)
the square lattice.\cite{Thereza_01,Paiva_05,Ehsan_Marcos_12} 
Results are presented for the Hubbard model with $U/w=1/2$, $1$, $3/2$, and $2$ at 
half filling, and for the Heisenberg model with $J/w = 1/18$ (honeycomb) and 
$J/w = 1/32$ (square). The NLCE site expansion for the Heisenberg model 
in the square lattice was carried out up to the $fifteenth$ order.\cite{Tang_12} 
Lines (symbols) depict results from NLCE (DQMC).}
\label{fig:Cv_vs_T}
\end{figure}

The specific heat of the Hubbard model in the honeycomb and square lattices are 
presented in Figs.~\ref{fig:Cv_vs_T}(a) and \ref{fig:Cv_vs_T}(b), respectively,
for different values of the coupling strength. The NLCE results in the square 
geometry were previously reported in Ref.~\onlinecite{Ehsan_Marcos_12}, while 
early DQMC results in the square and honeycomb geometries (admittedly less 
accurate than those reported here) can be found in Refs.~\onlinecite{Thereza_01} 
and \onlinecite{Paiva_05}, respectively. The complementarity of NLCE and DQMC 
is once again clear from these plots for both lattice geometries. 
For weak interactions (e.g., $U/w=1/2$), DQMC and NLCE results are in good 
agreement with each other down to the temperatures at which NLCE converges. 
However, DQMC can also access the very low-temperature regime that is not accessible
to NLCE. Because of this, DQMC can resolve the double-peak structure that is 
present in the square lattice for $U>0$, which is the result of the Mott insulating 
ground state with long-range AF order that occurs for any nonzero 
value of $U$. Such a structure is absent in the honeycomb lattice in the weakly 
interacting regime, where the ground state of the system lacks long-range 
AF order.\cite{Paiva_05}

For values of the interaction strength of the order of the bandwidth, we do find 
small deviations between NLCE and DQMC results at intermediate to low temperatures 
($T/w \lesssim0.1$ for both geometries), as (mainly) systematic errors due to the 
discretization of imaginary time start to affect the DQMC results.
Nevertheless, the relatively good agreement between the results from these 
two methods for $U=w$, especially for the honeycomb geometry down to 
$T/w\sim0.03$, indicates that systematic errors in DQMC are not 
playing an important role. For larger values of $U$ ($U/w=3/2$ and 2 in 
Fig.~\ref{fig:Cv_vs_T}), NLCE provides more accurate results down to lower
temperatures. In particular, from the NLCE data it becomes apparent that the 
position of the low-temperature peak moves to lower temperature (proportional to $1/U$) 
as the interaction strength increases. This is expected once the charge 
degrees of freedom become irrelevant and the system falls in a regime 
that can be described by the Heisenberg model. The specific heat predicted 
by the corresponding spin model in the honeycomb and square lattices are also depicted 
in Fig.~\ref{fig:Cv_vs_T}. They closely follow (but are not equal to) the
results obtained within the Hubbard model for the largest value of $U$ 
shown in those figures ($U/w=2$).

\begin{figure}[!t] 
\includegraphics[width=0.47\textwidth]{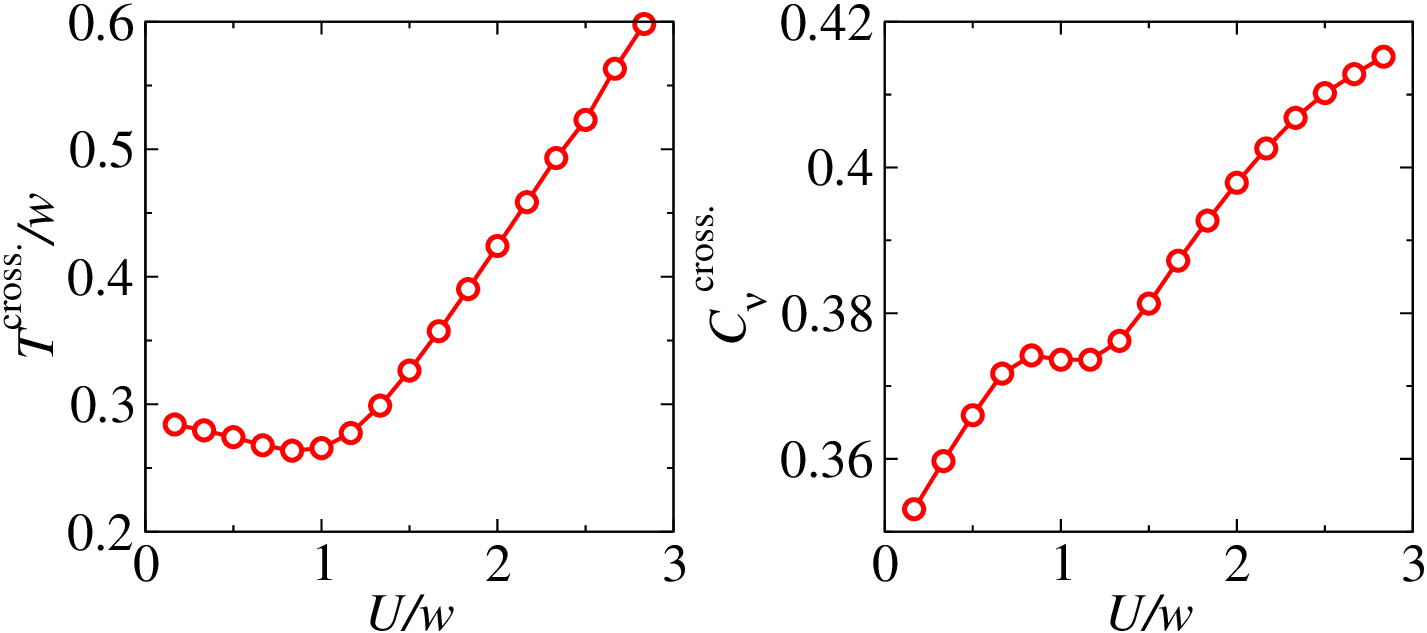}
\caption{(Color online) (a) Temperature and (b) $C_{v}$ at the
high-temperature crossing point between the specific heat curves for 
consecutive values of $U$ (see text), as obtained from NLCE for the honeycomb 
lattice.}
\label{fig:Cv_crossing}
\end{figure}

Similarly to what happens in the square lattice, in the weak-coupling 
regime in the honeycomb lattice the curves for specific heat vs temperature 
show what appears to be a unique crossing point at a high temperature for 
different interaction strengths.\cite{Paiva_05} This phenomenon has been 
extensively discussed for strongly-correlated models and 
materials.\cite{d_vollhardt_97} The high accuracy of the NLCE results 
at those temperatures allows us to make definite statements as to whether 
this is indeed a unique crossing point. In Fig.~\ref{fig:Cv_crossing}(a) 
and \ref{fig:Cv_crossing}(b), we plot (as a function of $U$) the high-$T$ 
crossing temperature ($T^\text{cross.}$) and the value of $C_v$ at the 
crossing ($C_v^\text{cross.}$) between curves of $C_v$ for $U$ 
and $U+w/6$. One can see that $T^\text{cross.}/w$ has a weak dependence 
on $U$ with a shallow minimum around $U/w=0.8$ before rising linearly 
with increasing the interaction at larger $U$. $C_v^\text{cross.}$ 
is also $U$ dependent. These results are qualitatively similar to 
those obtained for the square lattice in Ref.~\onlinecite{Ehsan_Marcos_12}, 
so that the features seen in $T^\text{cross.}$ and $C_v^\text{cross.}$ 
appear to be related to the onset of strong correlations in the system.

\subsection{Uniform Susceptibility and NN Spin Correlations}\label{suscep}

Another thermodynamic property of much experimental interest 
is the uniform susceptibility (per site)
\begin{equation}
 \chi = \frac{ \langle (\hat{S}^{z})^{2} \rangle -\langle \hat{S}^z \rangle^{2} }{NT},
\end{equation}
where $\hat{S^z}$ is the $z$ component of the spin operator. In solid state experiments,
$\chi$ is usually measured using SQUID magnetometers.

\begin{figure}[!t] 
\includegraphics[width=0.47\textwidth]{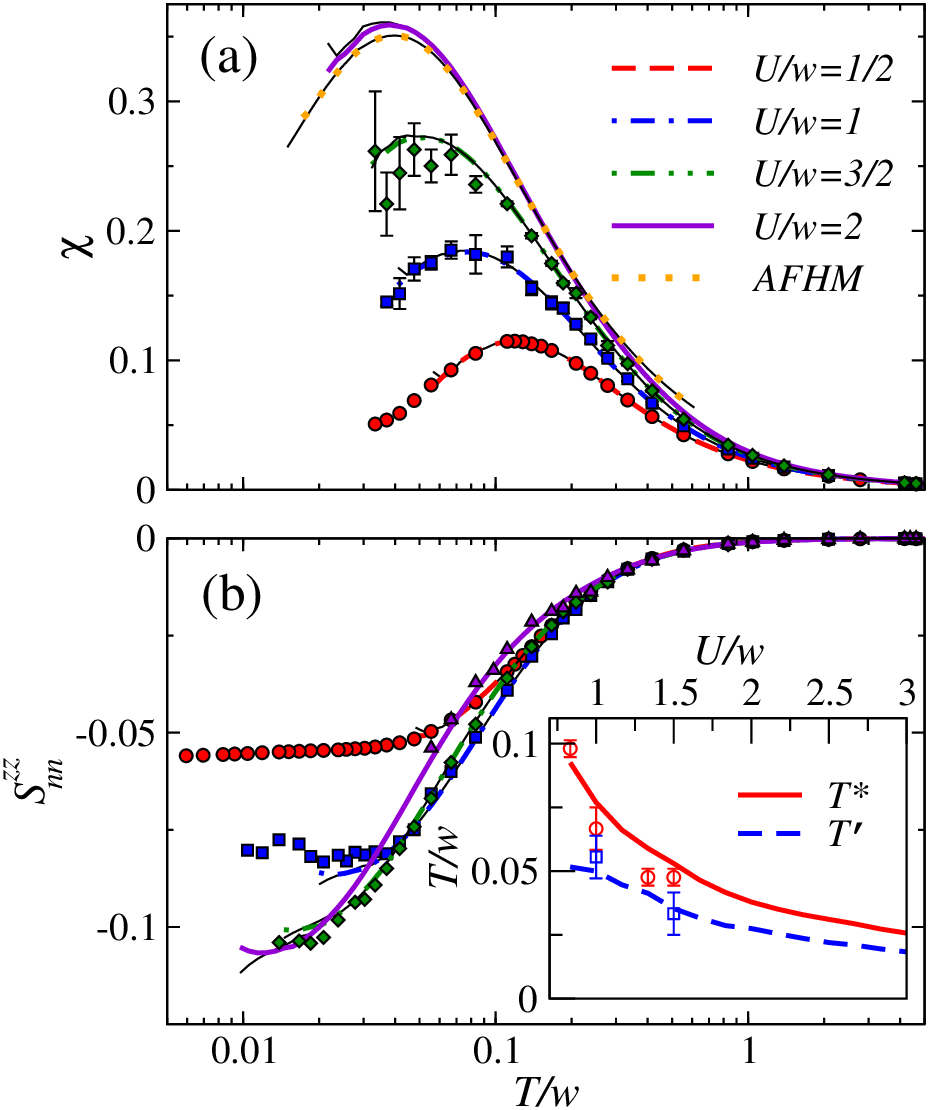}
\caption{(Color online) (a) Uniform spin susceptibility, and (b) NN spin correlations in
the honeycomb lattice at half filling vs $T$ for $U/w=1/2$, $1$, $3/2$, and $2$. 
In (a), we also include results for the Heisenberg model with $J/w=1/18$.
The inset in (b) shows $T^*$, which is the temperature at the maximum of the uniform 
susceptibility, and $T^\prime$, which is the temperature at which 
$\partial S^{zz}_{nn}/\partial T$ is maximal, as a function of $U$ 
(for $U > U_{c}$). Lines (symbols) depict results from NLCE (DQMC).}
\label{fig:chi_szz_vs_T}
\end{figure}

Results for the uniform susceptibility in the honeycomb lattice are presented in 
Fig.~\ref{fig:chi_szz_vs_T}(a) as a function of temperature for different interaction 
strengths. For $U/w\lesssim 1$, the DQMC results agree once again with those from NLCE 
down to the lowest convergence temperature in the NLCE. For $U/w>1$, an accurate DQMC 
calculation of this quantity at low temperatures becomes costly and we only show results 
for $U/w=1.5$, which exhibit large statistical errors at the lowest temperatures.

The plots for the uniform susceptibility in Fig.~\ref{fig:chi_szz_vs_T}(a) make apparent 
that there exists a peak for all values of $U$ whose location moves to lower temperatures 
monotonically with increasing $U$. This is unlike in the square lattice where such a peak 
first moves to higher temperatures with increasing $U$ (in the weakly interacting 
regime) before moving to lower temperatures for $U\gtrsim w$, following the AF characteristic 
temperature.\cite{Thereza_01,t_paiva_10,Ehsan_Marcos_11} Insight as to why the honeycomb 
lattice model behaves differently can be gained from the fact that its ground state 
is semimetallic in the weakly interacting regime with no long-range AF correlations. 
Therefore, the peak in the uniform susceptibility in that regime does not represent the 
onset of long-range AF correlations, which we know do not diverge in the ground state. 
In the strongly interacting regime, on the other hand, the peak temperature of $\chi$ (denoted 
by $T^*$) does characterize the onset of exponentially long AF correlations and decreases 
following the energy scale $J\propto 1/U$ as $U$ increases. In the inset in 
Fig.~\ref{fig:chi_szz_vs_T}(b), we plot $T^*$ as a function of $U$.

If it would be possible to measure the uniform susceptibility in experiments with 
ultracold gases, its downturn as the temperature is decreased could be used to 
identify the onset of antiferromagnetism. 
Since this is still not possible in current experimental setups, measuring NN spin 
correlations ($S^{zz}_{nn}$) can be considered a first step for probing the emergence 
of AF order in strongly-correlated lattice models such as the Hubbard model. 
This step has already been taken in experiments in the square lattice 
geometry.\cite{Trotzky_10,Greif_11,greif13} 

In our calculations, we define $S^{zz}_{nn}$ (per bond) as
\begin{equation}
S^{zz}_{nn}= \frac{1}{N {\pazocal Z}} 
\sum_{i,\langle j \rangle_i} \langle S^{z}_{i} S^{z}_{j} \rangle,
\end{equation} 
where ${\pazocal Z}$ is the coordination number (${\pazocal Z}=3$ for the honeycomb 
lattice), $\sum_{i,\langle j \rangle_i}$ means that we sum over all $j$'s that are nearest 
neighbors of $i$, and then sum over all $i$'s. In previous studies, we used NLCEs and 
DQMC to explore NN spin correlations at various interaction strengths in the square and 
honeycomb lattice geometries.\cite{Ehsan_Marcos_11,Tang_12_01} (Note that there is a 
factor 4 difference between the definition of $S^{zz}_{nn}$ here and in those references.) 
Surprisingly, we found that for a wide range of temperatures (entropies), accessible to 
current optical lattice experiments, these correlations are greater in the honeycomb 
than in the square lattice. This occurs even in regimes where the former has a semimetallic 
ground state while the latter has an AF Mott insulating one.\cite{Tang_12_01} 

Here, we are interested in identifying how to use those short range correlations to 
determine whether the system is in a regime with exponentially long 
AF correlations. In Fig.~\ref{fig:chi_szz_vs_T}(b), we show $S^{zz}_{nn}$, 
as a function of temperature in the honeycomb lattice and for different values of $U$.
They do not exhibit any sharp feature at $T^*$. For $U/w = 1/2$, which is below the critical 
value for the formation of the Mott insulator in the ground state of the honeycomb lattice 
model, $|S^{zz}_{nn}|$ increases but eventually saturates as $T$ decreases. Note that the 
minus sign indicates opposite NN (pseudo-)spins. For $U>U_{c}$, $|S^{zz}_{nn}|$ increases 
to values much larger than for $U<U_{c}$ with the region of fastest increase being pushed 
to lower temperatures as $U$ is increased. We have computed the derivative of $S^{zz}_{nn}$
with respect to temperature ($\partial S^{zz}_{nn}/\partial T$) and determined the temperature 
at which it is maximal. That temperature (denoted by $T'$) is shown in the inset in 
Fig.~\ref{fig:chi_szz_vs_T}(b). One can see there that $T^\prime$ is close to, but falls below, 
$T^*$. This implies that if $\partial S^{zz}_{nn}/\partial T$ is determined from experimental 
measurements and $T'$ is identified, then one will know (from the results here) that the 
system is in a regime with exponentially long AF correlations for $T\lesssim T'$.

\subsection{Staggered susceptibility and structure factor} 

In Sec.~\ref{suscep}, we discussed two indirect measurements that help locating
the onset of antiferromagnetism in the Hubbard model in the honeycomb lattice. 
Here we present results for two quantities that directly track the growth
of such correlations, namely, the staggered susceptibility and the AF
structure factor.

The staggered susceptibility (per site) is defined as the negative second derivative 
of the free energy (per site) with respect to a staggered field, $h$, added to the Hamiltonian:
\begin{equation}
 \chi^{\textrm{stg}} =\frac{T}{N}\frac{\partial^2 \ln Z }{\partial h^2}|_{h=0}.
\end{equation}
This derivative can be evaluated numerically within NLCE because one has 
full access to the partition function in the presence or absence of a 
staggered field. Using a Taylor expansion of $\ln Z$, and considering that 
the magnetization at $h=0$ is zero,
$\chi^{\textrm{stg}}$ can be rewritten as
\begin{equation}
 \chi^{\textrm{stg}} = \frac{2T}{N} \frac{\ln Z|_{h=\Delta h} -\ln Z|_{h=0}}{\Delta h^2},
\end{equation} 
where $\Delta h$ is very small. In the NLCE calculations, we have tested several values of 
$\Delta h$, which were orders of magnitude apart ($0.001 \lesssim \Delta h \lesssim 0.1$), 
to ensure that the value of $\chi^{\textrm{stg}}$ reported here is independent of the 
$\Delta h$ chosen. 

\begin{figure}[!t] 
\includegraphics[width=0.45\textwidth]{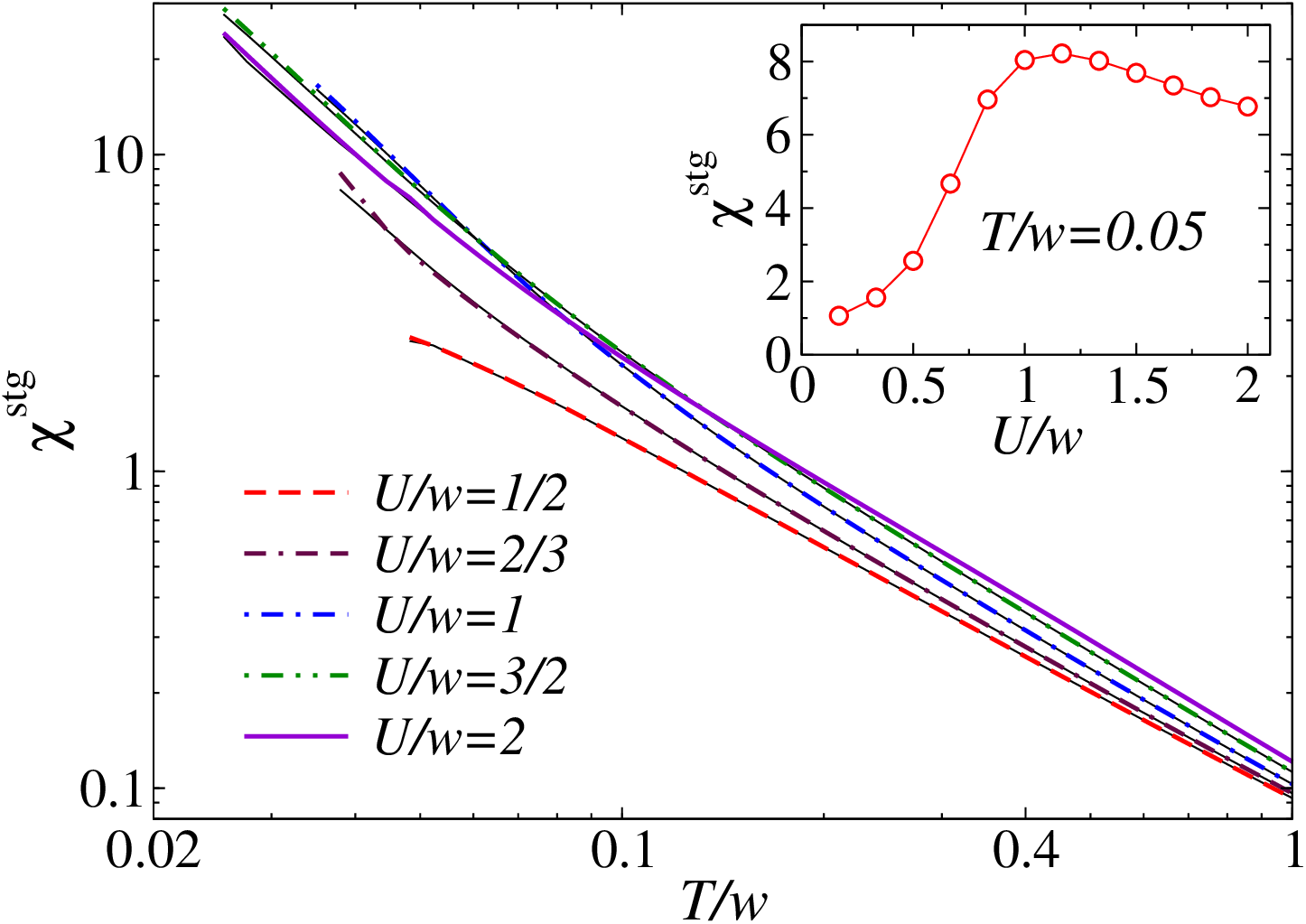}
\caption{(Color online) NLCE results for the staggered susceptibility as a function of 
temperature in the honeycomb lattice at half filling for $U/w$ from $1/2$ up to $2$. 
The inset shows $\chi^\textrm{stg}$ as a function of $U/w$ for $T/w=0.05$.}
\label{fig:stagger}
\end{figure}

In Fig.~\ref{fig:stagger}, we show results for the staggered susceptibility
as a function of temperature for various interaction strengths. In all cases,
one can see a fast growth of this quantity when lowering the temperature 
starting from high temperatures ($T\sim w$). However, when $T/w\lesssim 0.1$, the 
curves for $U>U_c$ depart from those for $U<U_c$, as the curvature of 
$\chi^\textrm{stg}$ changes sign by changing $U$. In fact, for temperatures 
lower than $T^*$ for $U>U_c$ (see the inset in Fig.~\ref{fig:chi_szz_vs_T}), 
$\chi^{\textrm{stg}}$ appears to be growing exponentially. 

As discussed before, $J$ in the effective Heisenberg model decreases as $U$ 
increases in the strongly interacting regime, which results in the onset of 
exponentially growing correlations to move to lower temperatures with 
increasing $U$ (for $U\gtrsim w$). This has a visible effect on 
$\chi^\textrm{stg}$ at the lowest temperatures we have access to. 
In Fig.~\ref{fig:stagger}, one can already see that after an initial
increase for $U\le w$, $\chi^\textrm{stg}$ decreases 
with increasing $U$ for $U>w$.
This is shown more clearly in the inset of Fig.~\ref{fig:stagger}, where we 
plot $\chi^\textrm{stg}$ vs $U/w$ at $T/w=0.05$. The maximal value of 
$\chi^\textrm{stg}$ at that temperature occurs in the vicinity 
of $U=w$.

The second quantity of interest is the AF structure factor, which is defined as
\begin{equation}
 S_\text{AF}=\frac{1}{N}\sum_{ij} \vartheta_{i,j}\left<S^{z}_{i} S^{z}_{j}\right>,
\end{equation}
where $\vartheta_{i,j}=-1$ if $i$ and $j$ belong to a different sublattice and 
$\vartheta_{i,j}=1$ if $i$ and $j$ belong to the same sublattice. Note that
the lack of long range order at any finite temperature implies that $S_\text{AF}$ 
is finite in the thermodynamic limit. This means that for the NLCE calculation
an ``additive'' structure factor is computed in finite clusters as 
$\sum_{ij} \vartheta_{i,j}\left<S^{z}_{i} S^{z}_{j}\right>$. This is done so that the 
intensive counterpart ($S_\text{AF}$) can be evaluated within NLCE in the 
thermodynamic limit. $S_\text{AF}$ can be measured in solid-state systems 
using neutron scattering, and, in ultracold atomic gases, it has been 
recently proposed that Bragg scattering can be used for that purpose.\cite{corcovilos10} 
It can also be calculated using DQMC.

\begin{figure}[!t] 
\includegraphics[width=0.45\textwidth]{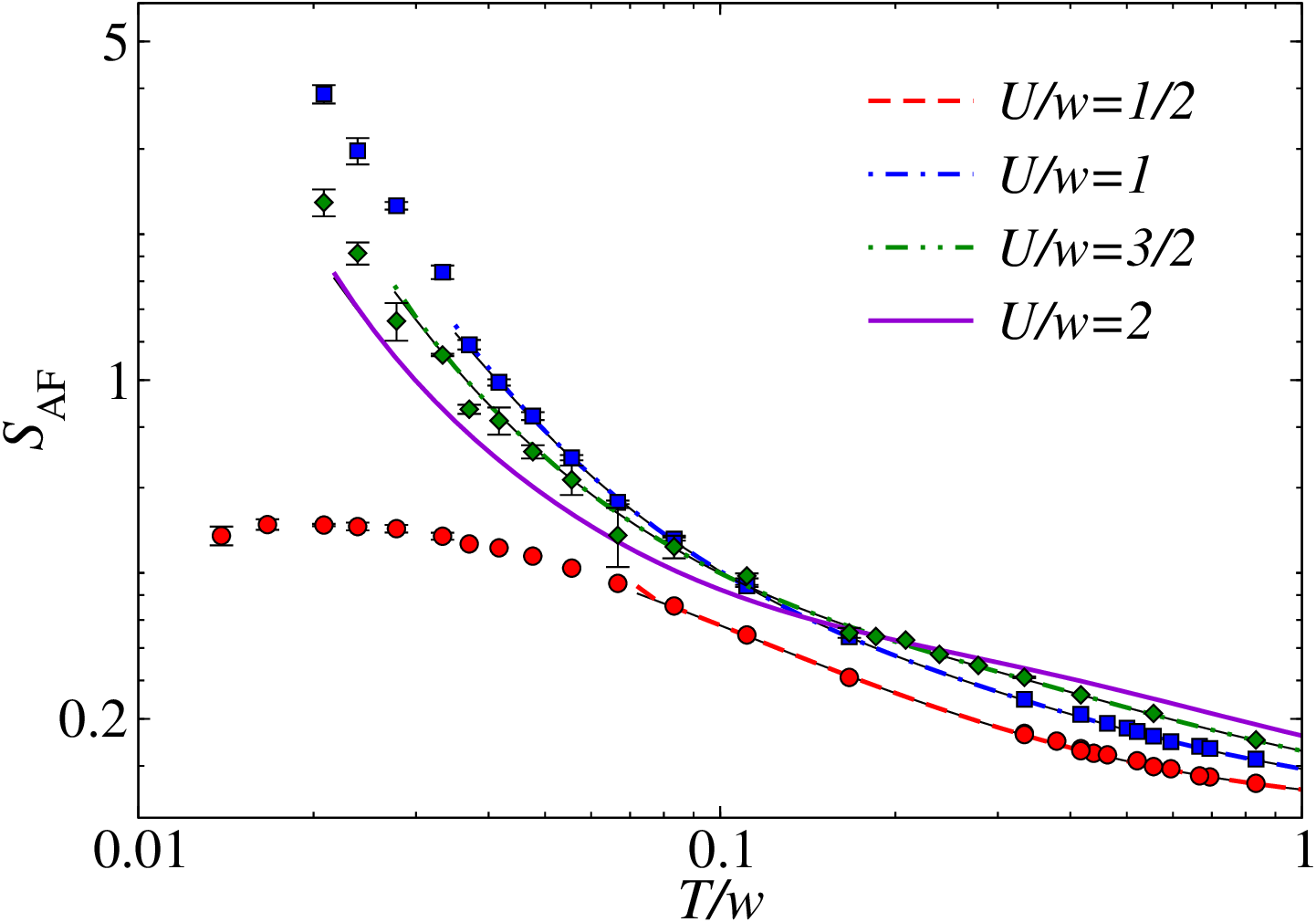}
\caption{(Color online) Antiferromagnetic structure factor as a function of 
temperature in the honeycomb lattice at half filling with $U/w = 1/2$, $1$, $3/2$, 
and $2$. Lines (symbols) correspond to NLCE (extrapolated DQMC) results.}
\label{fig:structure}
\end{figure}

In Fig.~\ref{fig:structure}, we plot $S_\text{AF}$ as a function of $T$ for the
honeycomb lattice at half filling for different values of $U/w$. We should note
that, in contrast to the quantities discussed in previous sections, 
$S_\text{AF}$ exhibits strong finite-size effects in the DQMC simulations 
for the temperatures of interest here. Because of this, the DQMC results shown 
in Fig.~\ref{fig:structure} are the outcome of an extrapolation to the thermodynamic 
limit considering five different system sizes (24, 54, 96, 150, and 216 sites). 
Closely related to the strong finite-size effects seen in the DQMC calculations, 
we note that NLCE results for $S_\text{AF}$ generally converge up 
to temperatures (for a given $U$) that are higher than those for the quantities 
discussed in previous sections.

It is apparent in Fig.~\ref{fig:structure} that the results for $S_\text{AF}$ 
as a function of temperature and interaction strength correlate with those for
$\chi^\textrm{stg}$ in Fig.~\ref{fig:stagger}. For $S_\text{AF}$, we show 
DQMC results that extend to much lower temperatures than those at which NLCE 
converges in the weak-coupling regime. They make evident that, for $U<U_c$ 
($U/w = 1/2$ in Fig.~\ref{fig:structure}) where the ground state is a semimetal, 
the structure factor increases very slowly with decreasing temperature below
$T/w=0.1$ before saturating and even decreasing at the lowest temperatures
accessible to DQMC. For strong interactions, on the other hand, the growth 
below $T=T^*$  is exponential, which is consistent with the presence of an 
AF ground state. 

\subsection{Double occupancy}

A quantity that draws great interest in experiments with ultracold gases in optical
lattices is the double occupancy $D=\langle\hat{n}_{\uparrow}\hat{n}_{\downarrow}\rangle$, 
which can be accurately measured and has been used to identify the presence
of a Mott insulator.\cite{jordens_strohmaier_08} In our recent Letter,\cite{Tang_12_01} 
we found that, in the half-filled Hubbard model on the honeycomb geometry, 
$D$ exhibits a more pronounced low-temperature rise with decreasing temperature than 
in the square lattice. This implies that, as discussed there and in what follows 
in Sec.~\ref{cooling}, adiabatic cooling is more efficient in the honeycomb lattice.

\begin{figure}[!t] 
\includegraphics[width=0.475\textwidth]{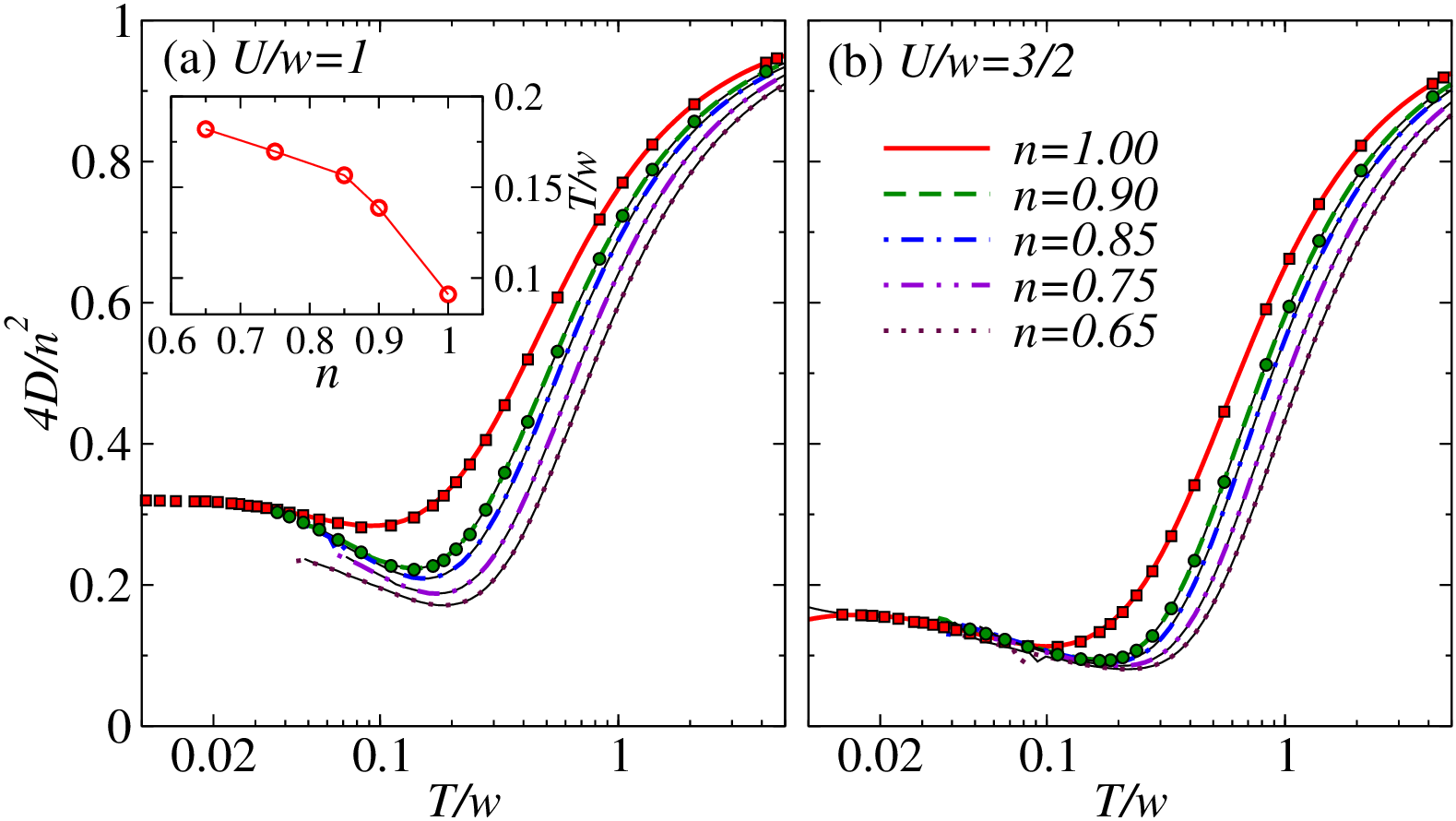}
\caption{(Color online) NLCE (lines) and DQMC (symbols) results for normalized double 
occupancy $4D/n^2$ vs $T$ at different densities for (a) $U/w=1$ and (b) $U/w=3/2$
on the honeycomb lattice. 
Inset of (a) shows $T/w$ at the minimum of $4D/n^2$ as a function of $n$.}
\label{fig:dob}
\end{figure}

Here we show what happens away from half filling. Figure~\ref{fig:dob} depicts the 
normalized double occupancy, namely, $D$ divided by its uncorrelated high-temperature 
value of $n^2/4$, for two different values of the interaction strength. 
In Fig.~\ref{fig:dob}, one can see that the region in temperature with an anomalous 
$dD/dT<0$ exists not only at half filling but also away from it. The low-$T$ upturn away 
from half filling, which has also been observed for the square lattice 
geometry,\cite{Mikelsons_09,Ehsan_Marcos_11} can be attributed to delocalization effects 
due to Fermi liquid behavior, especially in the weak to intermediate coupling regimes. 
Consistently with that picture, the low-$T$ upturn starts at higher temperatures as $n$ 
decreases from 1, as shown in the inset in Fig.~\ref{fig:dob}(a). This also explains the 
enhancement in the value of $|dD/dT|$ in the anomalous region as one dopes the system 
away from, but remains close to, half filling for $U/w=1$. At half filling, this phenomenon, 
and the resulting Pomeranchuk cooling mechanism, has been studied using 
several techniques.~\cite{Werner_05,a_dare_07,l_deleo_08} 

In the strong-coupling regime, the (less pronounced) upturn in $D$ at half filling is 
attributable to the increase in virtual hoppings to NN sites due to the enhancement of 
AF correlations.~\cite{e_gorelik_10} In that case, the similar upturn in $D$ away from 
half filling, and the agreement of the normalized double occupancies for a range of 
densities close to one at low temperatures [see Fig.~\ref{fig:dob}(b)], can be explained 
by the fact that AF correlations and ordering of the moments likely remain significant, 
even at small dopings when the interaction 
strength is large.~\cite{Ehsan_Marcos_11}

\subsection{Adiabatic cooling away from half filling} \label{cooling}

As mentioned before, the existence of an anomalous region with $dD/dT<0$ is of much 
interest as it implies that the system can be cooled adiabatically by increasing the 
interaction strength. This follows from the relation 
$\partial S/\partial U=-\partial D/\partial T$,\cite{Werner_05}
which indicates that the entropy $S$ increases with increasing $U$ at constant $T$, 
or $T$ decreases with increasing $U$ at constant $S$. In Ref.~\onlinecite{Tang_12_01}, 
we compared the constant-entropy curves in the $T$-$U$ plane between the honeycomb and 
square lattice at half filling, and showed that the adiabatic cooling is more efficient 
in the former lattice geometry. 

\begin{figure}[!b] 
\includegraphics[width=3.3in]{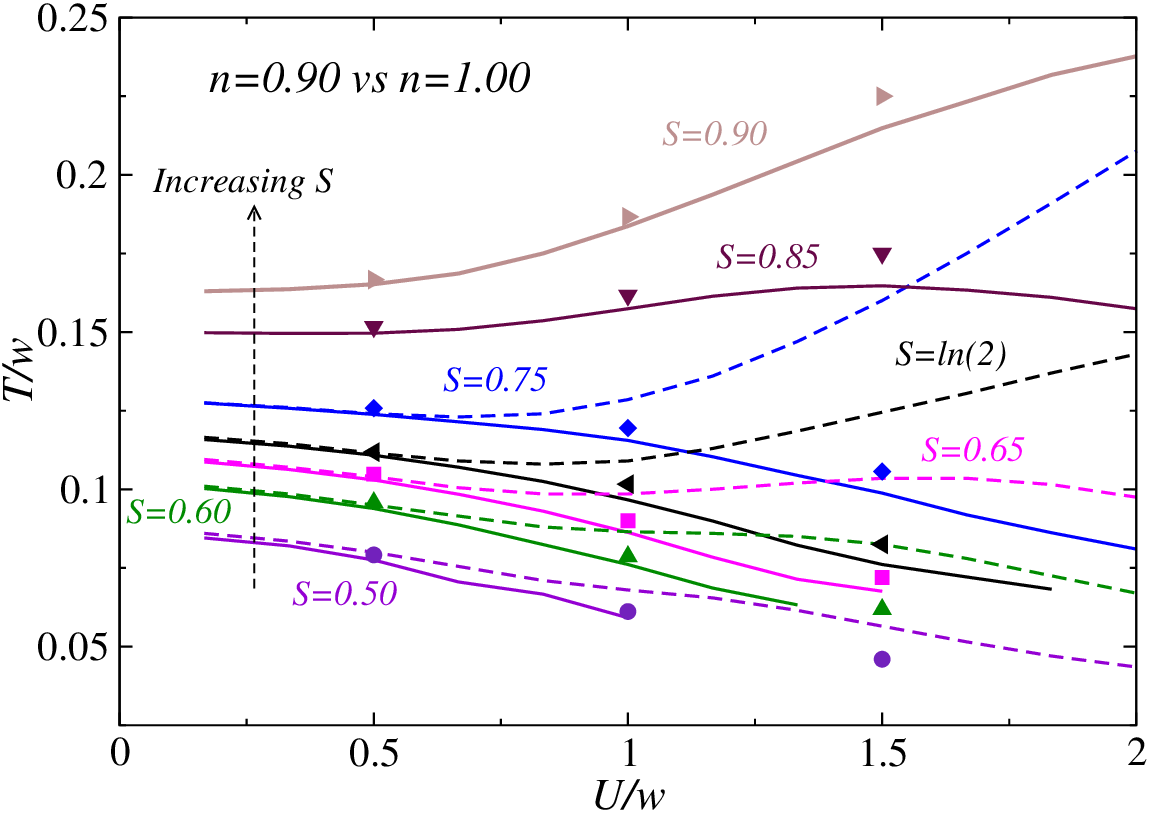}
\caption{(Color online) Isentropic curves for $T/w$ vs $U/w$ in the honeycomb lattice 
at density $n = 1.00$ (thin dashed lines) and $n = 0.90$ (thick solid lines) for constant 
entropies $S=0.50$, $0.60$, $0.65$, $ln(2)$, $0.75$, $0.85$, and $0.90$. Those results
were obtained using NLCE. For $n = 0.90$, we also present results from DQMC (symbols).}
\label{fig:cooling_away}
\end{figure}

Here, we present results for $n\ne1$, and explore how such a cooling mechanism changes
away from half filling relative to the half-filled case. In Fig.~\ref{fig:cooling_away}, 
we show the isentropic curves in the $T$-$U$ plane at $n=0.90$ (NLCE as thick solid lines 
and DQMC as symbols) and compare them with the ones at $n=1.00$ (NLCE as thin dashed lines).
It is worth noting the good agreement between NLCE and DQMC results for $n=0.90$, 
specially for $U/w<1$. For the ranges of $U/w$ and temperatures presented here 
the ``sign'' problem in the QMC calculations was small at this filling. 

Figure~\ref{fig:cooling_away} makes apparent what one could have
predicted from the results for the double occupancy in Fig.~\ref{fig:dob}
(from the more pronounced rise in $D$ by lowering temperature when 
$n=0.9$ than when $n=1$), namely, that adiabatic cooling by increasing 
$U$ is generally stronger away, but close to, half filling. For high entropies, 
e.g., $S = 0.75$ in Fig.~\ref{fig:cooling_away}, one can see that as $U$ increases 
in the strong-coupling regime the temperature for the system with $n=0.9$ decreases 
while it increases for the half-filled system. Similarly, at lower entropies, e.g., 
$S=0.6$ in Fig.~\ref{fig:cooling_away}, the system with $n=0.9$ can be cooled down 
to lower temperatures than with $n=1.00$. As we show in the following section, 
these findings have important implications for adiabatically cooling of systems 
confined in harmonic potentials, as is the case for experiments with cold gases 
in optical lattices.

\section{Trapped Systems}
\label{sec:trapped}

In order to emulate optical lattice experiments, we consider fermions trapped in a
harmonic potential by adding a term $\sum_{i\sigma}V\,r_{i}^{2}\,\hat{n}_{i\sigma}$ 
to the Hamiltonian~\eqref{eq:Hamiltonian}, where $V$ is the potential strength and 
$r_{i}$ is the distance of each site from the center of the trap. Since NLCE and 
DQMC results discussed in the previous sections are for homogeneous systems, we use 
the LDA to estimate thermodynamic properties of the 
confined system. For temperatures similar to those studied here, LDA was found to be 
a good approximation, at least for the square lattice geometry, in a DQMC study of 
the Hubbard model for harmonically trapped systems.\cite{Simone_Christopher_11}

We are interested in understanding how adiabatic cooling changes in the presence of a 
confining potential. In those systems, the density changes from its maximal value in the 
center of the trap to zero. Given the results in Fig.~\ref{fig:cooling_away}, which 
show an important dependence of the isentropic curves on the density, one needs to 
consider a fully trapped system to determine how the contributions from low and high 
densities add up in a trap to result in cooling (or heating) as the interaction strength is 
increased. To provide general results, which apply to systems with potentially 
different strengths of the confining potential and/or number of particles, it is 
convenient to utilize the characteristic density 
$\tilde{\rho}=N_p(2V/w)^{d/2}$,\cite{rigol_muramatsu_03,Rigol200433} 
where $N_p$ is the total number of particles in the trap and $d$ is the dimensionality. 

\begin{figure}[!t] 
\includegraphics[width=3.3in]{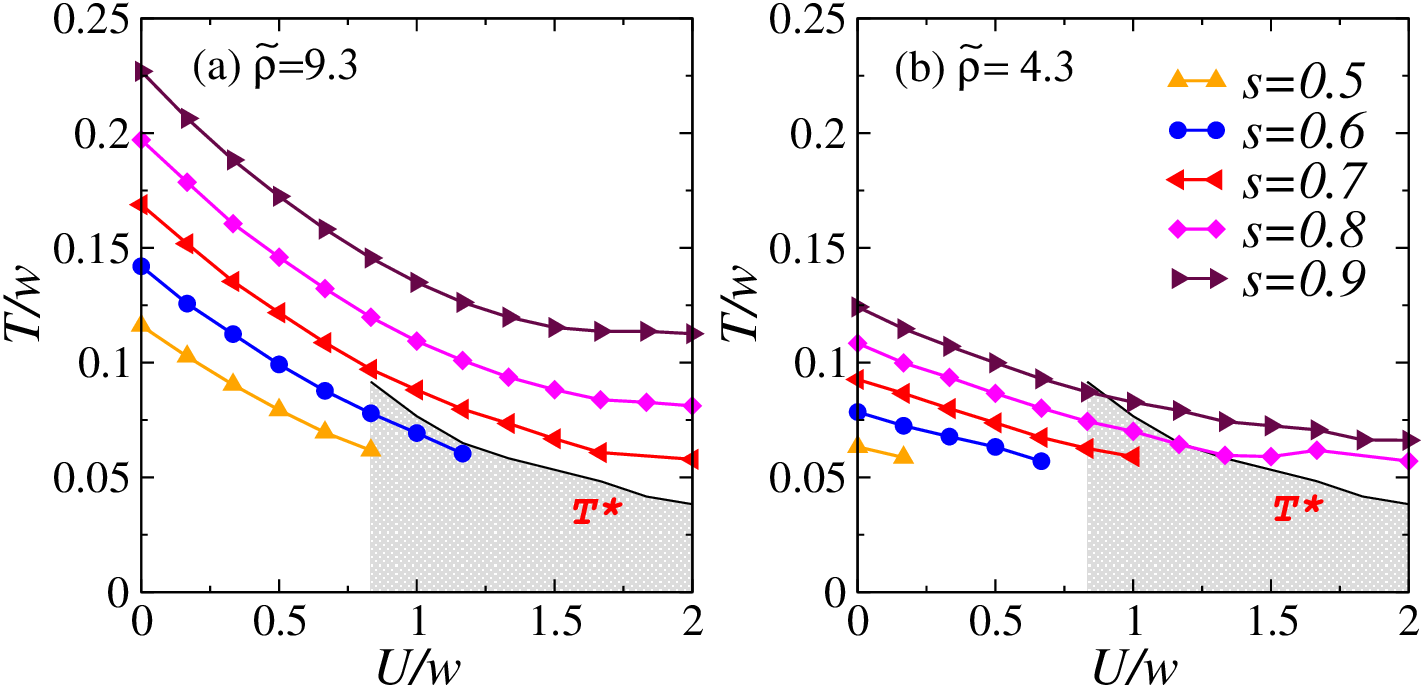}
\caption{(Color online) NLCE results for the isentropic curves of $T/w$ vs $U/w$ 
in trapped systems in the honeycomb lattice for constant average entropies per particle 
$s=0.5$, $0.6$, $0.7$, $0.8$, and $0.9$; and for (a) $\tilde{\rho}=9.3$ and (b) $\tilde{\rho}=4.3$. 
The shaded region emphasizes temperatures below $T^*$, for ($U>U_c$), in the half-filled 
homogeneous system [see the inset in Fig.~\ref{fig:chi_szz_vs_T}(b)].}
\label{fig:cooling_trap}
\end{figure}

In Fig.~\ref{fig:cooling_trap}, we show isentropic curves for $T/w$ vs $U/w$ in 
trapped systems with $\tilde{\rho}=9.3$ and $4.3$. Note that here, $s$ gives 
the average entropy per particle in the trap, which is to be compared to the entropy 
per site ($S$) of the homogeneous system only at half filling. For $U>U_c$, the region 
with $T<T^*$ in the latter system is depicted as a shaded area in 
Fig.~\ref{fig:cooling_trap}.  Figure~\ref{fig:cooling_trap} shows that even if $s$ is 
as large as $0.9$, the temperature in the trapped system decreases steadily as $U$ 
increases. This is to be contrasted with the results in Fig.~\ref{fig:cooling_away} 
where such a steady decrease of the temperature is not seen even if $S$ at $n=1$ is 
as low as 0.6. Furthermore, as shown in Fig.~\ref{fig:cooling_trap}(b), one can 
adiabatically drive the trapped system into a regime with $T<T^*$ starting from 
entropies per particle that are higher than $S=0.5$, required for the $n=1$
homogeneous case to achieve $T<T^*$.\cite{Tang_12_01}

\begin{figure}[!b] 
\includegraphics[width=3.3in]{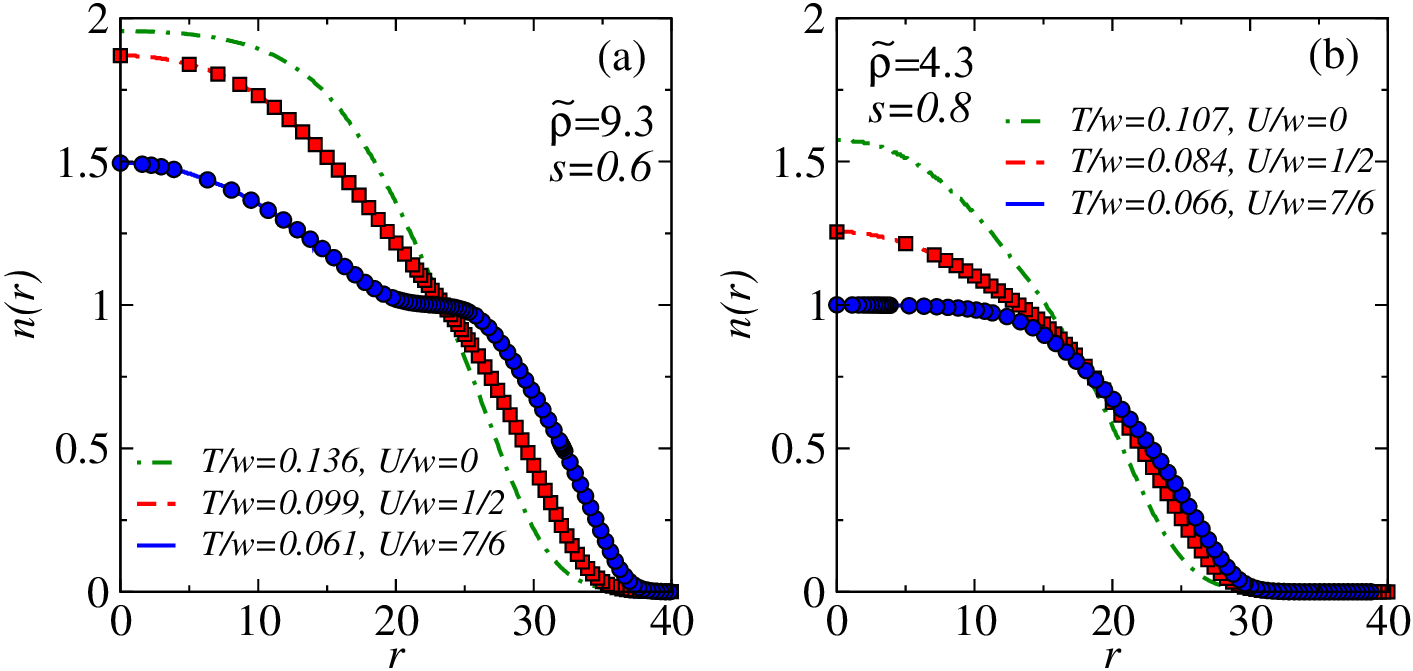}
\caption{(Color online) Local density for trapped fermions in the honeycomb lattice 
with (a) $\tilde{\rho}=9.3$ and $s=0.6$, and (b) $\tilde{\rho}=4.3$ and 
$s=0.8$. Results are presented for $U/w$ from $0$ to $7/6$. NLCE results are 
depicted as lines and DQMC results as symbols.}
\label{fig:density_trap}
\end{figure}

What remains to be checked is whether there is a Mott ($n=1$) region in the trapped
system when $T<T^*$. In Fig.~\ref{fig:density_trap}, we show density profiles 
corresponding to the same characteristic densities depicted in 
Fig.~\ref{fig:cooling_trap}, for selected values of the entropy per particle and 
$U/w$. Figure~\ref{fig:density_trap}(a) reports results for $\tilde{\rho}=9.3$ 
when $s=0.6$. This average entropy leads to a temperature $T<T^*$ when $U/w\simeq 1$, 
and we indeed find that a Mott shoulder develops in the system 
at the desired interaction strength (see the density profile for $U/w=7/6$). 

A preferable AF Mott plateau occupying the entire center of the trap
may then be obtained by reducing the characteristic density. This can be seen in 
Fig.~\ref{fig:cooling_trap}(b) for $\tilde{\rho}=4.3$ and $s=0.8$, where the density 
profile at $T/w=0.066$ and $U/w=7/6$ exhibits an extended plateau with $n=1$ in 
the center of the system. Note that, for this characteristic density and entropy per particle, 
the isentropic curve for $s=0.8$ exhibits a region in $U/w$ for which $T$ is 
below $T^*$ [Fig.~\ref{fig:cooling_trap}(b)]. For $\tilde{\rho}<4.3$ (not shown), 
we find that the density profiles for the highest values of $s$ whose temperature 
falls below $T^*$ have $n<1$ at $r=0$. This means that the highest average entropy 
per particle that can support exponentially long AF correlations in a Mott plateau 
in the center of the trap is roughly $0.8$, and this occurs for characteristic densities 
$\tilde{\rho}\simeq4.3$ [corresponding to the results in Fig.~\ref{fig:cooling_trap}(b)].

\begin{figure}[!t] 
\includegraphics[width=3.3in]{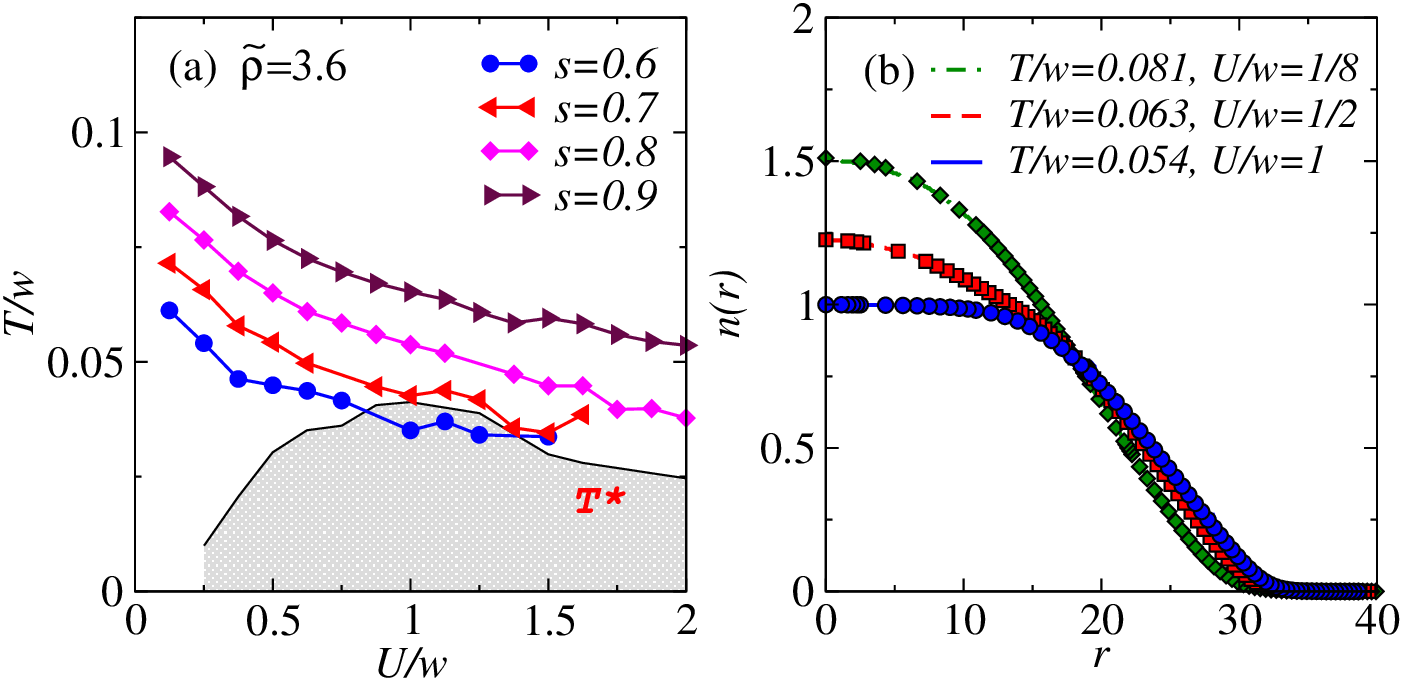}
\caption{(Color online) (a) NLCE results for the isentropic curves in the square lattice, 
similar to Fig.~\ref{fig:cooling_trap} but with $\tilde{\rho}=3.6$; (b) local density in 
the square lattice with $\tilde{\rho}=3.6$ and $s=0.8$. Lines and symbols are the same as 
in Fig.~\ref{fig:density_trap}.}
\label{fig:cooling_trap_sq}
\end{figure}

Finally, one may wonder how those results compare to what happens in the square
lattice. In Fig.~\ref{fig:cooling_trap_sq}(a), we show isentropic curves for a 
harmonically trapped system in the square lattice with $\tilde{\rho}=3.6$. This 
is about the lowest characteristic density that supports a Mott insulator with 
$n=1$ for $s\sim 0.8$. The density profiles corresponding to $s=0.8$ are presented 
in Fig.~\ref{fig:cooling_trap_sq}(b). (Those results are similar to the ones 
presented in Ref.~\onlinecite{Ehsan_Marcos_11}, but are extended to lower 
temperature using numerical resummations in the NLCE.) 
Figure~\ref{fig:cooling_trap_sq}(a) clearly shows that $s\simeq 0.6$ are required
to achieve temperatures below $T^*$ in this case, so that the Mott plateau 
seen in Fig.~\ref{fig:cooling_trap_sq}(b) for $U/w=1$ does not exhibit 
exponentially long correlations. This indicates that long-range AF correlations 
should be easier to observe in the honeycomb geometry than in the square one.

\section{summary}
\label{sec:summ}

In summary, we have utilized two complementary and unbiased computational 
techniques, NLCE and DQMC, to study finite-temperature thermodynamic properties 
of the Hubbard model in the honeycomb lattice. We have calculated experimentally 
relevant quantities, including the entropy, specific heat, uniform and staggered 
spin susceptibilities, NN spin correlations, and the double occupancy at and away 
from half filling. 

Among other things, we correlated the maximum in 
$\partial S^{zz}_{nn}/\partial T$ with the temperature $T^*$ at which 
exponentially long correlations set in the system, so that NN spin 
correlations could be used in experiments to (indirectly) 
probe antiferromagnetism. We have also shown that a low-temperature upturn 
in the double occupancy also occurs away from half filling, 
and that it is actually more prominent than the one at 
half filling. By comparing the isentropic curves for $T/w$ 
as a function of $U/w$ away from (but close to) half filling and those at half 
filling, we have shown that adiabatic cooling is more efficient in the former. 
We then have used the local density approximation to study adiabatic cooling 
in the presence of a confining harmonic potential, finding that in the 
trapped system such a process is indeed more efficient than in the homogeneous
one. This means that, in the presence of a harmonic trap, one can create Mott 
insulating domains with exponentially long AF correlations at 
average entropies per particle of $s\simeq0.8$, while in the homogeneous system 
one needs $s\lesssim 0.5$ to fall in such a regime.\cite{Tang_12_01} These upper
limits for the values of entropy are even higher than those predicted for the square 
lattice geometry.

\begin{acknowledgments}
This work was supported by NSF Grant No.~OCI-0904597 (B.T., E.K., and M.R.),
and by CNPq, FAPERJ and INCT on Quantum Information (T.P.).
\end{acknowledgments}


\begin{thebibliography}{51}
\expandafter\ifx\csname natexlab\endcsname\relax\def\natexlab#1{#1}\fi
\expandafter\ifx\csname bibnamefont\endcsname\relax
  \def\bibnamefont#1{#1}\fi
\expandafter\ifx\csname bibfnamefont\endcsname\relax
  \def\bibfnamefont#1{#1}\fi
\expandafter\ifx\csname citenamefont\endcsname\relax
  \def\citenamefont#1{#1}\fi
\expandafter\ifx\csname url\endcsname\relax
  \def\url#1{\texttt{#1}}\fi
\expandafter\ifx\csname urlprefix\endcsname\relax\def\urlprefix{URL }\fi
\providecommand{\bibinfo}[2]{#2}
\providecommand{\eprint}[2][]{\url{#2}}

\bibitem[{\citenamefont{Castro~Neto et~al.}(2009)\citenamefont{Castro~Neto,
  Guinea, Peres, Novoselov, and Geim}}]{castroneto09}
\bibinfo{author}{\bibfnamefont{A.~H.} \bibnamefont{Castro~Neto}},
  \bibinfo{author}{\bibfnamefont{F.}~\bibnamefont{Guinea}},
  \bibinfo{author}{\bibfnamefont{N.~M.~R.} \bibnamefont{Peres}},
  \bibinfo{author}{\bibfnamefont{K.~S.} \bibnamefont{Novoselov}},
  \bibnamefont{and} \bibinfo{author}{\bibfnamefont{A.~K.} \bibnamefont{Geim}},
  \bibinfo{journal}{Rev. Mod. Phys.} \textbf{\bibinfo{volume}{81}},
  \bibinfo{pages}{109} (\bibinfo{year}{2009}).

\bibitem[{\citenamefont{Meng et~al.}(2010)\citenamefont{Meng, Lang, Wessel,
  Assaad, and Muramatsu}}]{meng_lang_10}
\bibinfo{author}{\bibfnamefont{Z.}~\bibnamefont{Meng}},
  \bibinfo{author}{\bibfnamefont{T.}~\bibnamefont{Lang}},
  \bibinfo{author}{\bibfnamefont{S.}~\bibnamefont{Wessel}},
  \bibinfo{author}{\bibfnamefont{F.}~\bibnamefont{Assaad}}, \bibnamefont{and}
  \bibinfo{author}{\bibfnamefont{A.}~\bibnamefont{Muramatsu}},
  \bibinfo{journal}{Nature (London)} \textbf{\bibinfo{volume}{464}},
  \bibinfo{pages}{847} (\bibinfo{year}{2010}).

\bibitem[{\citenamefont{Wu et~al.}(2010)\citenamefont{Wu, Chen, Tao, Tong, and
  Liu}}]{w_wu_10}
\bibinfo{author}{\bibfnamefont{W.}~\bibnamefont{Wu}},
  \bibinfo{author}{\bibfnamefont{Y.-H.} \bibnamefont{Chen}},
  \bibinfo{author}{\bibfnamefont{H.-S.} \bibnamefont{Tao}},
  \bibinfo{author}{\bibfnamefont{N.-H.} \bibnamefont{Tong}}, \bibnamefont{and}
  \bibinfo{author}{\bibfnamefont{W.-M.} \bibnamefont{Liu}},
  \bibinfo{journal}{Phys. Rev. B} \textbf{\bibinfo{volume}{82}},
  \bibinfo{pages}{245102} (\bibinfo{year}{2010}).

\bibitem[{\citenamefont{Liebsch}(2011)}]{a_liebsch_11}
\bibinfo{author}{\bibfnamefont{A.}~\bibnamefont{Liebsch}},
  \bibinfo{journal}{Phys. Rev. B} \textbf{\bibinfo{volume}{83}},
  \bibinfo{pages}{035113} (\bibinfo{year}{2011}).

\bibitem[{\citenamefont{Ebrahimkhas}(2011)}]{m_ebrahimkhas_11}
\bibinfo{author}{\bibfnamefont{M.}~\bibnamefont{Ebrahimkhas}},
  \bibinfo{journal}{Phys. Lett. A} \textbf{\bibinfo{volume}{375}},
  \bibinfo{pages}{3223 } (\bibinfo{year}{2011}).

\bibitem[{\citenamefont{Ma et~al.}(2011)\citenamefont{Ma, Huang, Hu, and
  Lin}}]{t_ma_11}
\bibinfo{author}{\bibfnamefont{T.}~\bibnamefont{Ma}},
  \bibinfo{author}{\bibfnamefont{Z.}~\bibnamefont{Huang}},
  \bibinfo{author}{\bibfnamefont{F.}~\bibnamefont{Hu}}, \bibnamefont{and}
  \bibinfo{author}{\bibfnamefont{H.-Q.} \bibnamefont{Lin}},
  \bibinfo{journal}{Phys. Rev. B} \textbf{\bibinfo{volume}{84}},
  \bibinfo{pages}{121410} (\bibinfo{year}{2011}).

\bibitem[{\citenamefont{Wang et~al.}(2011)\citenamefont{Wang, Goerbig,
  Miniatura, and Gr\'emaud}}]{g_wang_11}
\bibinfo{author}{\bibfnamefont{G.}~\bibnamefont{Wang}},
  \bibinfo{author}{\bibfnamefont{M.~O.} \bibnamefont{Goerbig}},
  \bibinfo{author}{\bibfnamefont{C.}~\bibnamefont{Miniatura}},
  \bibnamefont{and}
  \bibinfo{author}{\bibfnamefont{B.}~\bibnamefont{Gr\'emaud}},
  \bibinfo{journal}{Europh. Lett.)} \textbf{\bibinfo{volume}{95}},
  \bibinfo{pages}{47013} (\bibinfo{year}{2011}).

\bibitem[{\citenamefont{Sun and Kou}(2011)}]{g_sun_11}
\bibinfo{author}{\bibfnamefont{G.-Y.} \bibnamefont{Sun}} \bibnamefont{and}
  \bibinfo{author}{\bibfnamefont{S.-P.} \bibnamefont{Kou}},
  \bibinfo{journal}{J. Phys.: Condens. Matter} \textbf{\bibinfo{volume}{23}},
  \bibinfo{pages}{045603} (\bibinfo{year}{2011}).

\bibitem[{\citenamefont{Tang et~al.}(2012)\citenamefont{Tang, Paiva, Khatami,
  and Rigol}}]{Tang_12_01}
\bibinfo{author}{\bibfnamefont{B.}~\bibnamefont{Tang}},
  \bibinfo{author}{\bibfnamefont{T.}~\bibnamefont{Paiva}},
  \bibinfo{author}{\bibfnamefont{E.}~\bibnamefont{Khatami}}, \bibnamefont{and}
  \bibinfo{author}{\bibfnamefont{M.}~\bibnamefont{Rigol}},
  \bibinfo{journal}{Phys. Rev. Lett.} \textbf{\bibinfo{volume}{109}},
  \bibinfo{pages}{205301} (\bibinfo{year}{2012}).

\bibitem[{\citenamefont{He and Lu}(2012)}]{r_he_12}
\bibinfo{author}{\bibfnamefont{R.-Q.} \bibnamefont{He}} \bibnamefont{and}
  \bibinfo{author}{\bibfnamefont{Z.-Y.} \bibnamefont{Lu}},
  \bibinfo{journal}{Phys. Rev. B} \textbf{\bibinfo{volume}{86}},
  \bibinfo{pages}{045105} (\bibinfo{year}{2012}).

\bibitem[{\citenamefont{Chang and Scalettar}(2012)}]{c_chang_12}
\bibinfo{author}{\bibfnamefont{C.-C.} \bibnamefont{Chang}} \bibnamefont{and}
  \bibinfo{author}{\bibfnamefont{R.~T.} \bibnamefont{Scalettar}},
  \bibinfo{journal}{Phys. Rev. Lett.} \textbf{\bibinfo{volume}{109}},
  \bibinfo{pages}{026404} (\bibinfo{year}{2012}).

\bibitem[{\citenamefont{Sorella et~al.}(2012)\citenamefont{Sorella, Otsuka, and
  Yunoki}}]{Sandro_Yuichi_12}
\bibinfo{author}{\bibfnamefont{S.}~\bibnamefont{Sorella}},
  \bibinfo{author}{\bibfnamefont{Y.}~\bibnamefont{Otsuka}}, \bibnamefont{and}
  \bibinfo{author}{\bibfnamefont{S.}~\bibnamefont{Yunoki}},
  \bibinfo{journal}{Sci. Rep.} \textbf{\bibinfo{volume}{2}},
  \bibinfo{pages}{992} (\bibinfo{year}{2012}).

\bibitem[{\citenamefont{Wu et~al.}(2013)\citenamefont{Wu, Scherer, Honerkamp,
  and Le~Hur}}]{w_wu_13}
\bibinfo{author}{\bibfnamefont{W.}~\bibnamefont{Wu}},
  \bibinfo{author}{\bibfnamefont{M.~M.} \bibnamefont{Scherer}},
  \bibinfo{author}{\bibfnamefont{C.}~\bibnamefont{Honerkamp}},
  \bibnamefont{and} \bibinfo{author}{\bibfnamefont{K.}~\bibnamefont{Le~Hur}},
  \bibinfo{journal}{Phys. Rev. B} \textbf{\bibinfo{volume}{87}},
  \bibinfo{pages}{094521} (\bibinfo{year}{2013}).

\bibitem[{\citenamefont{Hassan and S\'en\'echal}(2013)}]{s_hassan_13}
\bibinfo{author}{\bibfnamefont{S.~R.} \bibnamefont{Hassan}} \bibnamefont{and}
  \bibinfo{author}{\bibfnamefont{D.}~\bibnamefont{S\'en\'echal}},
  \bibinfo{journal}{Phys. Rev. Lett.} \textbf{\bibinfo{volume}{110}},
  \bibinfo{pages}{096402} (\bibinfo{year}{2013}).

\bibitem[{\citenamefont{Assaad and Herbut}(unpublished)}]{assaad_herbut}
\bibinfo{author}{\bibfnamefont{F.~F.} \bibnamefont{Assaad}} \bibnamefont{and}
  \bibinfo{author}{\bibfnamefont{I.~F.} \bibnamefont{Herbut}},
Phys. Rev. X {\bf 3}, 031010 (2013).

\bibitem[{\citenamefont{Sorella and Tosatti}(1992)}]{s_sorella_92}
\bibinfo{author}{\bibfnamefont{S.}~\bibnamefont{Sorella}} \bibnamefont{and}
  \bibinfo{author}{\bibfnamefont{E.}~\bibnamefont{Tosatti}},
  \bibinfo{journal}{EPL (Europhysics Letters)} \textbf{\bibinfo{volume}{19}},
  \bibinfo{pages}{699} (\bibinfo{year}{1992}).

\bibitem[{\citenamefont{Paiva et~al.}(2005)\citenamefont{Paiva, Scalettar,
  Zheng, Singh, and Oitmaa}}]{Paiva_05}
\bibinfo{author}{\bibfnamefont{T.}~\bibnamefont{Paiva}},
  \bibinfo{author}{\bibfnamefont{R.~T.} \bibnamefont{Scalettar}},
  \bibinfo{author}{\bibfnamefont{W.}~\bibnamefont{Zheng}},
  \bibinfo{author}{\bibfnamefont{R.~R.~P.} \bibnamefont{Singh}},
  \bibnamefont{and} \bibinfo{author}{\bibfnamefont{J.}~\bibnamefont{Oitmaa}},
  \bibinfo{journal}{Phys. Rev. B} \textbf{\bibinfo{volume}{72}},
  \bibinfo{pages}{085123} (\bibinfo{year}{2005}).

\bibitem[{\citenamefont{Santoro et~al.}(1993)\citenamefont{Santoro, Airoldi,
  Sorella, and Tosatti}}]{g_santoro_93}
\bibinfo{author}{\bibfnamefont{G.}~\bibnamefont{Santoro}},
  \bibinfo{author}{\bibfnamefont{M.}~\bibnamefont{Airoldi}},
  \bibinfo{author}{\bibfnamefont{S.}~\bibnamefont{Sorella}}, \bibnamefont{and}
  \bibinfo{author}{\bibfnamefont{E.}~\bibnamefont{Tosatti}},
  \bibinfo{journal}{Phys. Rev. B} \textbf{\bibinfo{volume}{47}},
  \bibinfo{pages}{16216} (\bibinfo{year}{1993}).

\bibitem[{\citenamefont{Jafari}(2009)}]{s_jafari_09}
\bibinfo{author}{\bibfnamefont{S.~A.} \bibnamefont{Jafari}},
  \bibinfo{journal}{Eur. Phys. J. B} \textbf{\bibinfo{volume}{68}},
  \bibinfo{pages}{537} (\bibinfo{year}{2009}).

\bibitem[{\citenamefont{Le}(2013)}]{d_le_13}
\bibinfo{author}{\bibfnamefont{D.~A.} \bibnamefont{Le}}, \bibinfo{journal}{Mod.
  Phys. Lett. B} \textbf{\bibinfo{volume}{27}}, \bibinfo{pages}{1350046}
  (\bibinfo{year}{2013}).

\bibitem[{\citenamefont{Singha et~al.}(2011)\citenamefont{Singha, Gibertini,
  Karmakar, Yuan, Polini, Vignale, Katsnelson, Pinczuk, Pfeiffer, West
  et~al.}}]{honeycom_sci}
\bibinfo{author}{\bibfnamefont{A.}~\bibnamefont{Singha}},
  \bibinfo{author}{\bibfnamefont{M.}~\bibnamefont{Gibertini}},
  \bibinfo{author}{\bibfnamefont{B.}~\bibnamefont{Karmakar}},
  \bibinfo{author}{\bibfnamefont{S.}~\bibnamefont{Yuan}},
  \bibinfo{author}{\bibfnamefont{M.}~\bibnamefont{Polini}},
  \bibinfo{author}{\bibfnamefont{G.}~\bibnamefont{Vignale}},
  \bibinfo{author}{\bibfnamefont{M.~I.} \bibnamefont{Katsnelson}},
  \bibinfo{author}{\bibfnamefont{A.}~\bibnamefont{Pinczuk}},
  \bibinfo{author}{\bibfnamefont{L.~N.} \bibnamefont{Pfeiffer}},
  \bibinfo{author}{\bibfnamefont{K.~W.} \bibnamefont{West}},
  \bibnamefont{{\it et~al.}}, \bibinfo{journal}{Science}
  \textbf{\bibinfo{volume}{332}}, \bibinfo{pages}{1176} (\bibinfo{year}{2011}).

\bibitem[{\citenamefont{Tarruell et~al.}(2012)\citenamefont{Tarruell, Greif,
  Uehlinger, Jotzu, and Esslinger}}]{honeycom_nat1}
\bibinfo{author}{\bibfnamefont{L.}~\bibnamefont{Tarruell}},
  \bibinfo{author}{\bibfnamefont{D.}~\bibnamefont{Greif}},
  \bibinfo{author}{\bibfnamefont{T.}~\bibnamefont{Uehlinger}},
  \bibinfo{author}{\bibfnamefont{G.}~\bibnamefont{Jotzu}}, \bibnamefont{and}
  \bibinfo{author}{\bibfnamefont{T.}~\bibnamefont{Esslinger}},
  \bibinfo{journal}{Nature (London)} \textbf{\bibinfo{volume}{483}},
  \bibinfo{pages}{302} (\bibinfo{year}{2012}).

\bibitem[{\citenamefont{Gomes et~al.}(2012)\citenamefont{Gomes, Mar, Ko,
  Guinea, and Manoharan}}]{honeycom_nat2}
\bibinfo{author}{\bibfnamefont{K.~K.} \bibnamefont{Gomes}},
  \bibinfo{author}{\bibfnamefont{W.}~\bibnamefont{Mar}},
  \bibinfo{author}{\bibfnamefont{W.}~\bibnamefont{Ko}},
  \bibinfo{author}{\bibfnamefont{F.}~\bibnamefont{Guinea}}, \bibnamefont{and}
  \bibinfo{author}{\bibfnamefont{H.~C.} \bibnamefont{Manoharan}},
  \bibinfo{journal}{Nature (London)} \textbf{\bibinfo{volume}{483}},
  \bibinfo{pages}{306} (\bibinfo{year}{2012}).

\bibitem[{\citenamefont{Esslinger}(2010)}]{esslinger_review_10}
\bibinfo{author}{\bibfnamefont{T.}~\bibnamefont{Esslinger}},
  \bibinfo{journal}{Annu. Rev. Condens. Matter Phys.}
  \textbf{\bibinfo{volume}{1}}, \bibinfo{pages}{129} (\bibinfo{year}{2010}).

\bibitem[{\citenamefont{J\"ordens et~al.}(2008)\citenamefont{J\"ordens,
  Strohmaier, G\"unter, Moritz, and Esslinger}}]{jordens_strohmaier_08}
\bibinfo{author}{\bibfnamefont{R.}~\bibnamefont{J\"ordens}},
  \bibinfo{author}{\bibfnamefont{N.}~\bibnamefont{Strohmaier}},
  \bibinfo{author}{\bibfnamefont{K.}~\bibnamefont{G\"unter}},
  \bibinfo{author}{\bibfnamefont{H.}~\bibnamefont{Moritz}}, \bibnamefont{and}
  \bibinfo{author}{\bibfnamefont{T.}~\bibnamefont{Esslinger}},
  \bibinfo{journal}{Nature (London)} \textbf{\bibinfo{volume}{455}},
  \bibinfo{pages}{204} (\bibinfo{year}{2008}).

\bibitem[{\citenamefont{Rigol et~al.}(2006)\citenamefont{Rigol, Bryant, and
  Singh}}]{Marcos_06}
\bibinfo{author}{\bibfnamefont{M.}~\bibnamefont{Rigol}},
  \bibinfo{author}{\bibfnamefont{T.}~\bibnamefont{Bryant}}, \bibnamefont{and}
  \bibinfo{author}{\bibfnamefont{R.~R.~P.} \bibnamefont{Singh}},
  \bibinfo{journal}{Phys. Rev. Lett.} \textbf{\bibinfo{volume}{97}},
  \bibinfo{pages}{187202} (\bibinfo{year}{2006}).

\bibitem[{\citenamefont{Rigol et~al.}(2007{\natexlab{a}})\citenamefont{Rigol,
  Bryant, and Singh}}]{Marcos_07_01}
\bibinfo{author}{\bibfnamefont{M.}~\bibnamefont{Rigol}},
  \bibinfo{author}{\bibfnamefont{T.}~\bibnamefont{Bryant}}, \bibnamefont{and}
  \bibinfo{author}{\bibfnamefont{R.~R.~P.} \bibnamefont{Singh}},
  \bibinfo{journal}{Phys. Rev. E} \textbf{\bibinfo{volume}{75}},
  \bibinfo{pages}{061118} (\bibinfo{year}{2007}{\natexlab{a}}).

\bibitem[{\citenamefont{Rigol et~al.}(2007{\natexlab{b}})\citenamefont{Rigol,
  Bryant, and Singh}}]{Marcos_07_02}
\bibinfo{author}{\bibfnamefont{M.}~\bibnamefont{Rigol}},
  \bibinfo{author}{\bibfnamefont{T.}~\bibnamefont{Bryant}}, \bibnamefont{and}
  \bibinfo{author}{\bibfnamefont{R.~R.~P.} \bibnamefont{Singh}},
  \bibinfo{journal}{Phys. Rev. E} \textbf{\bibinfo{volume}{75}},
  \bibinfo{pages}{061119} (\bibinfo{year}{2007}{\natexlab{b}}).

\bibitem[{\citenamefont{Scalapino and Sugar}(1981)}]{Scalapino_81}
\bibinfo{author}{\bibfnamefont{D.~J.} \bibnamefont{Scalapino}}
  \bibnamefont{and} \bibinfo{author}{\bibfnamefont{R.~L.} \bibnamefont{Sugar}},
  \bibinfo{journal}{Phys. Rev. Lett.} \textbf{\bibinfo{volume}{46}},
  \bibinfo{pages}{519} (\bibinfo{year}{1981}).

\bibitem[{\citenamefont{Tang et~al.}(2013)\citenamefont{Tang, Khatami, and
  Rigol}}]{Tang_12}
\bibinfo{author}{\bibfnamefont{B.}~\bibnamefont{Tang}},
  \bibinfo{author}{\bibfnamefont{E.}~\bibnamefont{Khatami}}, \bibnamefont{and}
  \bibinfo{author}{\bibfnamefont{M.}~\bibnamefont{Rigol}},
  \bibinfo{journal}{Comput. Phys. Commun.} \textbf{\bibinfo{volume}{184}},
  \bibinfo{pages}{557 } (\bibinfo{year}{2013}).

\bibitem[{\citenamefont{Hirsch}(1983)}]{hirsch_83}
\bibinfo{author}{\bibfnamefont{J.~E.} \bibnamefont{Hirsch}},
  \bibinfo{journal}{Phys. Rev. B} \textbf{\bibinfo{volume}{28}},
  \bibinfo{pages}{4059} (\bibinfo{year}{1983}).

\bibitem[{\citenamefont{Khatami and Rigol}(2011)}]{Ehsan_Marcos_11}
\bibinfo{author}{\bibfnamefont{E.}~\bibnamefont{Khatami}} \bibnamefont{and}
  \bibinfo{author}{\bibfnamefont{M.}~\bibnamefont{Rigol}},
  \bibinfo{journal}{Phys. Rev. A} \textbf{\bibinfo{volume}{84}},
  \bibinfo{pages}{053611} (\bibinfo{year}{2011}).

\bibitem[{\citenamefont{Loh et~al.}(1990)\citenamefont{Loh, Gubernatis,
  Scalettar, White, Scalapino, and Sugar}}]{sign1}
\bibinfo{author}{\bibfnamefont{E.~Y.} \bibnamefont{Loh}},
  \bibinfo{author}{\bibfnamefont{J.~E.} \bibnamefont{Gubernatis}},
  \bibinfo{author}{\bibfnamefont{R.~T.} \bibnamefont{Scalettar}},
  \bibinfo{author}{\bibfnamefont{S.~R.} \bibnamefont{White}},
  \bibinfo{author}{\bibfnamefont{D.~J.} \bibnamefont{Scalapino}},
  \bibnamefont{and} \bibinfo{author}{\bibfnamefont{R.~L.} \bibnamefont{Sugar}},
  \bibinfo{journal}{Phys. Rev. B} \textbf{\bibinfo{volume}{41}},
  \bibinfo{pages}{9301} (\bibinfo{year}{1990}).

\bibitem[{\citenamefont{Batrouni and Scalettar}(1990)}]{sign2}
\bibinfo{author}{\bibfnamefont{G.~G.} \bibnamefont{Batrouni}} \bibnamefont{and}
  \bibinfo{author}{\bibfnamefont{R.~T.} \bibnamefont{Scalettar}},
  \bibinfo{journal}{Phys. Rev. B} \textbf{\bibinfo{volume}{42}},
  \bibinfo{pages}{2282} (\bibinfo{year}{1990}).

\bibitem[{\citenamefont{Werner et~al.}(2005)\citenamefont{Werner, Parcollet,
  Georges, and Hassan}}]{Werner_05}
\bibinfo{author}{\bibfnamefont{F.}~\bibnamefont{Werner}},
  \bibinfo{author}{\bibfnamefont{O.}~\bibnamefont{Parcollet}},
  \bibinfo{author}{\bibfnamefont{A.}~\bibnamefont{Georges}}, \bibnamefont{and}
  \bibinfo{author}{\bibfnamefont{S.~R.} \bibnamefont{Hassan}},
  \bibinfo{journal}{Phys. Rev. Lett.} \textbf{\bibinfo{volume}{95}},
  \bibinfo{pages}{056401} (\bibinfo{year}{2005}).

\bibitem[{\citenamefont{Bon\ifmmode~\check{c}\else \v{c}\fi{}a and
  Prelov\ifmmode~\check{s}\else \v{s}\fi{}ek}(2003)}]{Bon_03}
\bibinfo{author}{\bibfnamefont{J.}~\bibnamefont{Bon\ifmmode~\check{c}\else
  \v{c}\fi{}a}} \bibnamefont{and}
  \bibinfo{author}{\bibfnamefont{P.}~\bibnamefont{Prelov\ifmmode~\check{s}\else
  \v{s}\fi{}ek}}, \bibinfo{journal}{Phys. Rev. B}
  \textbf{\bibinfo{volume}{67}}, \bibinfo{pages}{085103}
  (\bibinfo{year}{2003}).

\bibitem[{\citenamefont{Mikelsons et~al.}(2009)\citenamefont{Mikelsons,
  Khatami, Galanakis, Macridin, Moreno, and Jarrell}}]{Mikelsons_09}
\bibinfo{author}{\bibfnamefont{K.}~\bibnamefont{Mikelsons}},
  \bibinfo{author}{\bibfnamefont{E.}~\bibnamefont{Khatami}},
  \bibinfo{author}{\bibfnamefont{D.}~\bibnamefont{Galanakis}},
  \bibinfo{author}{\bibfnamefont{A.}~\bibnamefont{Macridin}},
  \bibinfo{author}{\bibfnamefont{J.}~\bibnamefont{Moreno}}, \bibnamefont{and}
  \bibinfo{author}{\bibfnamefont{M.}~\bibnamefont{Jarrell}},
  \bibinfo{journal}{Phys. Rev. B} \textbf{\bibinfo{volume}{80}},
  \bibinfo{pages}{140505} (\bibinfo{year}{2009}).

\bibitem[{\citenamefont{Khatami and Rigol}(2012)}]{Ehsan_Marcos_12}
\bibinfo{author}{\bibfnamefont{E.}~\bibnamefont{Khatami}} \bibnamefont{and}
  \bibinfo{author}{\bibfnamefont{M.}~\bibnamefont{Rigol}},
  \bibinfo{journal}{Phys. Rev. A} \textbf{\bibinfo{volume}{86}},
  \bibinfo{pages}{023633} (\bibinfo{year}{2012}).

\bibitem[{\citenamefont{Paiva et~al.}(2001)\citenamefont{Paiva, Scalettar,
  Huscroft, and McMahan}}]{Thereza_01}
\bibinfo{author}{\bibfnamefont{T.}~\bibnamefont{Paiva}},
  \bibinfo{author}{\bibfnamefont{R.~T.} \bibnamefont{Scalettar}},
  \bibinfo{author}{\bibfnamefont{C.}~\bibnamefont{Huscroft}}, \bibnamefont{and}
  \bibinfo{author}{\bibfnamefont{A.~K.} \bibnamefont{McMahan}},
  \bibinfo{journal}{Phys. Rev. B} \textbf{\bibinfo{volume}{63}},
  \bibinfo{pages}{125116} (\bibinfo{year}{2001}).

\bibitem[{\citenamefont{Vollhardt}(1997)}]{d_vollhardt_97}
\bibinfo{author}{\bibfnamefont{D.}~\bibnamefont{Vollhardt}},
  \bibinfo{journal}{Phys. Rev. Lett.} \textbf{\bibinfo{volume}{78}},
  \bibinfo{pages}{1307} (\bibinfo{year}{1997}).

\bibitem[{\citenamefont{Paiva et~al.}(2010)\citenamefont{Paiva, Scalettar,
  Randeria, and Trivedi}}]{t_paiva_10}
\bibinfo{author}{\bibfnamefont{T.}~\bibnamefont{Paiva}},
  \bibinfo{author}{\bibfnamefont{R.}~\bibnamefont{Scalettar}},
  \bibinfo{author}{\bibfnamefont{M.}~\bibnamefont{Randeria}}, \bibnamefont{and}
  \bibinfo{author}{\bibfnamefont{N.}~\bibnamefont{Trivedi}},
  \bibinfo{journal}{Phys. Rev. Lett.} \textbf{\bibinfo{volume}{104}},
  \bibinfo{pages}{066406} (\bibinfo{year}{2010}).

\bibitem[{\citenamefont{Trotzky et~al.}(2010)\citenamefont{Trotzky, Chen,
  Schnorrberger, Cheinet, and Bloch}}]{Trotzky_10}
\bibinfo{author}{\bibfnamefont{S.}~\bibnamefont{Trotzky}},
  \bibinfo{author}{\bibfnamefont{Y.-A.} \bibnamefont{Chen}},
  \bibinfo{author}{\bibfnamefont{U.}~\bibnamefont{Schnorrberger}},
  \bibinfo{author}{\bibfnamefont{P.}~\bibnamefont{Cheinet}}, \bibnamefont{and}
  \bibinfo{author}{\bibfnamefont{I.}~\bibnamefont{Bloch}},
  \bibinfo{journal}{Phys. Rev. Lett.} \textbf{\bibinfo{volume}{105}},
  \bibinfo{pages}{265303} (\bibinfo{year}{2010}).

\bibitem[{\citenamefont{Greif et~al.}(2011)\citenamefont{Greif, Tarruell,
  Uehlinger, J\"ordens, and Esslinger}}]{Greif_11}
\bibinfo{author}{\bibfnamefont{D.}~\bibnamefont{Greif}},
  \bibinfo{author}{\bibfnamefont{L.}~\bibnamefont{Tarruell}},
  \bibinfo{author}{\bibfnamefont{T.}~\bibnamefont{Uehlinger}},
  \bibinfo{author}{\bibfnamefont{R.}~\bibnamefont{J\"ordens}},
  \bibnamefont{and}
  \bibinfo{author}{\bibfnamefont{T.}~\bibnamefont{Esslinger}},
  \bibinfo{journal}{Phys. Rev. Lett.} \textbf{\bibinfo{volume}{106}},
  \bibinfo{pages}{145302} (\bibinfo{year}{2011}).

\bibitem[{\citenamefont{Greif et~al.}(2013)\citenamefont{Greif, Uehinger,
  Jotzu, Tarruell, and Esslinger}}]{greif13}
\bibinfo{author}{\bibfnamefont{D.}~\bibnamefont{Greif}},
  \bibinfo{author}{\bibfnamefont{T.}~\bibnamefont{Uehinger}},
  \bibinfo{author}{\bibfnamefont{G.}~\bibnamefont{Jotzu}},
  \bibinfo{author}{\bibfnamefont{L.}~\bibnamefont{Tarruell}}, \bibnamefont{and}
  \bibinfo{author}{\bibfnamefont{T.}~\bibnamefont{Esslinger}},
  \bibinfo{journal}{Science} \textbf{\bibinfo{volume}{340}},
  \bibinfo{pages}{1307} (\bibinfo{year}{2013}).

\bibitem[{\citenamefont{Corcovilos et~al.}(2010)\citenamefont{Corcovilos, Baur,
  Hitchcock, Mueller, and Hulet}}]{corcovilos10}
\bibinfo{author}{\bibfnamefont{T.~A.} \bibnamefont{Corcovilos}},
  \bibinfo{author}{\bibfnamefont{S.~K.} \bibnamefont{Baur}},
  \bibinfo{author}{\bibfnamefont{J.~M.} \bibnamefont{Hitchcock}},
  \bibinfo{author}{\bibfnamefont{E.~J.} \bibnamefont{Mueller}},
  \bibnamefont{and} \bibinfo{author}{\bibfnamefont{R.~G.} \bibnamefont{Hulet}},
  \bibinfo{journal}{Phys. Rev. A} \textbf{\bibinfo{volume}{81}},
  \bibinfo{pages}{013415} (\bibinfo{year}{2010}).

\bibitem[{\citenamefont{Dar\'e et~al.}(2007)\citenamefont{Dar\'e, Raymond,
  Albinet, and Tremblay}}]{a_dare_07}
\bibinfo{author}{\bibfnamefont{A.-M.} \bibnamefont{Dar\'e}},
  \bibinfo{author}{\bibfnamefont{L.}~\bibnamefont{Raymond}},
  \bibinfo{author}{\bibfnamefont{G.}~\bibnamefont{Albinet}}, \bibnamefont{and}
  \bibinfo{author}{\bibfnamefont{A.-M.~S.} \bibnamefont{Tremblay}},
  \bibinfo{journal}{Phys. Rev. B} \textbf{\bibinfo{volume}{76}},
  \bibinfo{pages}{064402} (\bibinfo{year}{2007}).

\bibitem[{\citenamefont{De~Leo et~al.}(2008)\citenamefont{De~Leo, Kollath,
  Georges, Ferrero, and Parcollet}}]{l_deleo_08}
\bibinfo{author}{\bibfnamefont{L.}~\bibnamefont{De~Leo}},
  \bibinfo{author}{\bibfnamefont{C.}~\bibnamefont{Kollath}},
  \bibinfo{author}{\bibfnamefont{A.}~\bibnamefont{Georges}},
  \bibinfo{author}{\bibfnamefont{M.}~\bibnamefont{Ferrero}}, \bibnamefont{and}
  \bibinfo{author}{\bibfnamefont{O.}~\bibnamefont{Parcollet}},
  \bibinfo{journal}{Phys. Rev. Lett.} \textbf{\bibinfo{volume}{101}},
  \bibinfo{pages}{210403} (\bibinfo{year}{2008}).

\bibitem[{\citenamefont{Gorelik et~al.}(2010)\citenamefont{Gorelik, Titvinidze,
  Hofstetter, Snoek, and Bl\"umer}}]{e_gorelik_10}
\bibinfo{author}{\bibfnamefont{E.~V.} \bibnamefont{Gorelik}},
  \bibinfo{author}{\bibfnamefont{I.}~\bibnamefont{Titvinidze}},
  \bibinfo{author}{\bibfnamefont{W.}~\bibnamefont{Hofstetter}},
  \bibinfo{author}{\bibfnamefont{M.}~\bibnamefont{Snoek}}, \bibnamefont{and}
  \bibinfo{author}{\bibfnamefont{N.}~\bibnamefont{Bl\"umer}},
  \bibinfo{journal}{Phys. Rev. Lett.} \textbf{\bibinfo{volume}{105}},
  \bibinfo{pages}{065301} (\bibinfo{year}{2010}).

\bibitem[{\citenamefont{Chiesa et~al.}(2011)\citenamefont{Chiesa, Varney,
  Rigol, and Scalettar}}]{Simone_Christopher_11}
\bibinfo{author}{\bibfnamefont{S.}~\bibnamefont{Chiesa}},
  \bibinfo{author}{\bibfnamefont{C.~N.} \bibnamefont{Varney}},
  \bibinfo{author}{\bibfnamefont{M.}~\bibnamefont{Rigol}}, \bibnamefont{and}
  \bibinfo{author}{\bibfnamefont{R.~T.} \bibnamefont{Scalettar}},
  \bibinfo{journal}{Phys. Rev. Lett.} \textbf{\bibinfo{volume}{106}},
  \bibinfo{pages}{035301} (\bibinfo{year}{2011}).

\bibitem[{\citenamefont{Rigol et~al.}(2003)\citenamefont{Rigol, Muramatsu,
  Batrouni, and Scalettar}}]{rigol_muramatsu_03}
\bibinfo{author}{\bibfnamefont{M.}~\bibnamefont{Rigol}},
  \bibinfo{author}{\bibfnamefont{A.}~\bibnamefont{Muramatsu}},
  \bibinfo{author}{\bibfnamefont{G.~G.} \bibnamefont{Batrouni}},
  \bibnamefont{and} \bibinfo{author}{\bibfnamefont{R.~T.}
  \bibnamefont{Scalettar}}, \bibinfo{journal}{Phys. Rev. Lett.}
  \textbf{\bibinfo{volume}{91}}, \bibinfo{pages}{130403}
  (\bibinfo{year}{2003}).

\bibitem[{\citenamefont{Rigol and Muramatsu}(2004)}]{Rigol200433}
\bibinfo{author}{\bibfnamefont{M.}~\bibnamefont{Rigol}} \bibnamefont{and}
  \bibinfo{author}{\bibfnamefont{A.}~\bibnamefont{Muramatsu}},
  \bibinfo{journal}{Opt. Commun.} \textbf{\bibinfo{volume}{243}},
  \bibinfo{pages}{33 } (\bibinfo{year}{2004}).

\end{thebibliography}

\end{document}